\documentclass[12pt]{article}
\usepackage{arxiv}
\usepackage{amsmath}
\usepackage{bm}

\usepackage{newtxtext,newtxmath}

\usepackage{float}

\usepackage[utf8]{inputenc}

\usepackage{lineno}
\usepackage{pdfpages}
\usepackage{siunitx}

\usepackage{enumitem}

\usepackage{graphicx}
\usepackage[export]{adjustbox}
\usepackage{needspace}
\usepackage{tabularx}

\usepackage{wrapfig}
\usepackage{pdflscape}
\usepackage[referable]{threeparttablex}

\usepackage[labelfont = bf, justification = justified, singlelinecheck = false, font = small, figurename = FIGURE, tablename = TABLE]{caption}

\usepackage{subcaption}

\usepackage{color,soul}
\sethlcolor{lime}

\usepackage[titletoc,toc,title]{appendix}

\usepackage{setspace}

\usepackage[english]{babel}

\usepackage{titlesec}

\titleformat*{\section}{\normalfont\bfseries}
\titleformat*{\subsection}{\normalfont\bfseries}
\titleformat*{\subsubsection}{\normalfont\itshape}

\usepackage{longtable}
\usepackage{multirow}
\usepackage{booktabs} 
\usepackage{afterpage}
\usepackage{relsize}
\usepackage{cprotect}
\usepackage{xcolor}

\usepackage{rotating}

\newcounter{tablenote}[table]

\usepackage{natbib}

\usepackage{fancyhdr}

\usepackage{array}
\newcolumntype{P}[1]{>{\centering\arraybackslash}p{#1}}

\allowdisplaybreaks

\usepackage[colorlinks,citecolor=blue,urlcolor=blue,bookmarks=false,hypertexnames=true]{hyperref}

\AtBeginDocument{%
}

\usepackage{lipsum}

\usepackage{autonum}

\begin{document}
    
    \doublespacing
    \frenchspacing
    
    {\Large\bf Outlier-robust copula regression for bivariate continuous proportions: an application to cushion plant vitality}

    \vskip 0cm

    {\bfseries
        Divan A. Burger$^{1,2}$,
        Janet van Niekerk$^{3,4}$,
        Peter C. le Roux$^{5}$,
        Morgan J. Raath-Krüger$^{6}$
    }

    \vspace{0.5\baselineskip}

    {\itshape\small
    \setlength{\tabcolsep}{0pt}
    \begin{ThreePartTable}
        \begin{longtable}{@{}p{1.5em}@{\hspace{0pt}}p{\dimexpr\textwidth-1.5em}@{}}
            $^{1}$ & Syneos Health, Bloemfontein, Free State, South Africa \\[2pt]
            $^{2}$ & Dept. of Mathematical Statistics and Actuarial Science, University of the Free State, Bloemfontein, Free State, South Africa \\[2pt]
            $^{3}$ & CEMSE Division, King Abdullah University of Science and Technology, Thuwal, Saudi Arabia \\[2pt]
            $^{4}$ & Dept. of Statistics, University of Pretoria, Pretoria, South Africa \\[2pt]
            $^{5}$ & Dept. of Plant and Soil Sciences, University of Pretoria, Pretoria, South Africa \\[2pt]
            $^{6}$ & Dept. of Botany and Plant Biotechnology, University of Johannesburg, Johannesburg, South Africa \\
        \end{longtable}
    \end{ThreePartTable}}

    %\noindent\rule{\textwidth}{1pt}
    %Divan Burger, Syneos Health, Bloemfontein, South Africa 9301. \\
    %Email: \verb|divanaburger@gmail.com|
    
    \section*{Abstract}
    
    Continuous proportions measured on the same experimental unit often pose two challenges: interior outliers that inflate variance beyond the beta ceiling and residual dependence that invalidates independent-margin models. We introduce a Bayesian copula modeling approach that combines rectangular-beta margins, which temper interior outliers by reallocating mass from the peak to a uniform component, with a single-parameter copula to capture concordance. Gaussian, Gumbel, and Clayton copula families are fitted, and log marginal likelihoods are obtained via bridge sampling to guide model selection. Applied to a 13-year survey (2003-2016) of \textit{Azorella selago} cushion plants on sub-Antarctic Marion Island, the copula models outperform independence baselines in explaining percent dead stem cover. Accounting for between-year dependence uncovers a positive west-slope effect and weakens the cushion size effect. Simulation results show negligible bias and near-nominal 95\% highest posterior density coverage across a range of tail weight and dependence scenarios, confirming good frequentist properties. The method integrates readily with \texttt{JAGS} and provides a robust default for paired proportion data in ecology and other disciplines where bounded outcomes and occasional outliers coincide.
    
    \vskip 0.5cm

    \setlength{\tabcolsep}{0.0cm}
    \begin{ThreePartTable}
        \begin{longtable}{p {2.3cm} p {14.2cm}}
            {\bf Keywords}: & Bayesian; Copula regression; Outlier robustness; Plant ecology; Rectangular-beta distribution.
        \end{longtable}
    \end{ThreePartTable}
    
    \setcounter{table}{0}
    
    \newpage
    
    \section{Introduction}

    Proportions on the open interval $\left(0,1\right)$ are common in ecology, for example, leaf area consumed by herbivores, ground cover of a species, biomass fractions, and the proportion of dead plant tissue \citep{defries2000new, poorter2012biomass}. Beta regression is a standard starting point: a logit link models the mean and a precision parameter absorbs extra variation \citep{Ferrari2004}. Field data, however, often violate beta assumptions because (i) a few extreme interior values can create residual variance patterns incompatible with beta margins, whose variance is bounded by 0.25, and (ii) many ecological surveys remeasure the same units over time, for example permanent-plot resurveys of percent cover \citep{jandt2022resurveygermany}, so any model must capture dependence while respecting the unit interval.
    
    Two strategies are common for inducing dependence between repeated proportions. The first augments independent beta margins with a normally distributed unit-level random effect so that, conditional on this effect, observations are independent. The second links beta margins through the beta copula of \citet{Joe2014Dependence}, yielding a parsimonious dependence model. Both strategies inherit the classical beta distribution's sensitivity to interior outliers, which motivates more robust margins.
    
    Heavy-tailed margins provide one remedy, including the rectangular-beta mixture \citep{Bayes2012}, the flexible beta mixture \citep{Migliorati2018FlexBeta}, and the rectangular-Kumaraswamy mixture \citep{castro2024robust}. As an alternative, the Johnson-$t$ distribution \citep{burger2023robust} achieves bounded heavy tails through a degrees of freedom parameter rather than a mixture component.
    
    We merge robust margins with copula-based dependence. Each margin follows the rectangular-beta distribution; its weight parameter shifts probability mass from the peak to a uniform component and buffers interior as well as mild boundary outliers. A one-parameter copula then introduces a single dependence term, which we parameterize via Kendall's $\tau$. We fit three copula families, Gaussian (symmetric), Gumbel (upper tail), and Clayton (lower tail). For comparison, we also fit two independence baselines, namely a rectangular-beta model and a conventional beta model, giving five candidate specifications in total.
    
    We illustrate the model with a 13-year photographic survey of the cushion plant \textit{Azorella selago} on sub-Antarctic Marion Island. Dead stem cover was recorded on individual cushions in 2003 and again in 2016 \citep{raath2023long}. Cushions are nested within eight plots, so we include a plot-level random intercept to absorb unmeasured site effects, while the copula captures between-year dependence. The model is fitted in a Bayesian framework, and bridge sampling is employed to estimate the log marginal likelihood (LML) when comparing the Gaussian, Gumbel, and Clayton copulas.
    
    Our contributions are robust rectangular-beta margins, a copula regression for paired proportions parameterized on the Kendall $\tau$ scale, and a practical workflow for model comparison and dependence assessment that, in application, changes the conclusions once between-year dependence is modeled.
    
    The rest of the paper is organized as follows. \autoref{sec:dataset} introduces the \textit{Azorella} data. \autoref{sec:rectbetacop} defines the bivariate rectangular-beta copula, and \autoref{sec:mixedmodel} embeds it in a mixed-effects framework. Bayesian inference is outlined in \autoref{sec:inference}, with model comparison via LMLs in \autoref{sec:modelcomp}. \autoref{sec:diag} presents residual and dependence diagnostics. The model is applied in \autoref{sec:application}. \autoref{sec:simulation} summarizes a supporting simulation study, and \autoref{sec:discussion} concludes with the main findings and outlines possible extensions.

    \section{Cushion plant dataset}\label{sec:dataset}

    \textit{Azorella selago} is a long-lived, hemispherical cushion shrub that dominates mid- and high-elevation vegetation on sub-Antarctic Marion Island. Its compact, prostrate growth form provides a substrate that has more moderate thermal and moisture conditions \citep[e.g.,][]{Badano2006}, creating favorable microhabitats for grasses, other vascular plants, mosses, and liverworts. Cushion vitality is assessed by the percentage of surface stems that are dead; a higher proportion signals poorer health and a reduced capacity to form such microhabitats.

    The dataset analyzed here comes from the photographic monitoring survey of \citet{raath2023long}. In both 2003 and 2016, each cushion plant in the monitoring plots was photographed from an approximately fixed height, its outline was traced, and the percent cover of dead stems was measured. Photographs that were of poor quality or corresponded to cushions that had died between surveys were excluded.

    The survey spanned three elevation bands (low, $\approx 200$~m above sea level; mid, $\approx 400$~m; and high, $\approx 600$~m) and two slope aspects (east and west), yielding twelve monitoring plots. For this exemplar analysis, we exclude the four low-elevation sites, retaining the eight plots from the mid- and high-elevation sites.

    Let $k=1,\dots,8$ index plots and $i=1,\dots,n$ index cushions, with $n=308$ and $k\left(i\right)$ the plot containing cushion $i$. For cushion $i$, we observe a pair of dead stem proportions $\bm{Y}_i=\left(Y_{1i}, Y_{2i}\right)^{\top}$, measured in 2003 and 2016, respectively; subscripts $1$ and $2$ refer to the years. Two plot indicators encode elevation $\left(\text{AltMid}_{k\left(i\right)}\right)$ and aspect $\left(\text{AspWest}_{k\left(i\right)}\right)$. Each year, we also recorded three covariates: percent cover of the grass \textit{Agrostis magellanica} (a dominant perennial and the most common vascular plant growing on \textit{A.~selago}); log-transformed cushion area; and percent cover of all other vascular or non-vascular plants. These responses will later be modeled with rectangular-beta margins linked by a one-parameter copula.

    Cushions within the same plot experience similar microclimatic and edaphic conditions; therefore, we include a plot-specific random intercept for each survey year to account for this shared heterogeneity. The random intercepts are assumed to be Gaussian and independent across years so that $\mathrm{b}_{jk} \sim \mathcal{N}\left(0,\sigma_{j}^{2}\right)$, and each cushion inherits its plot’s effect via $\mathrm{b}_{ji}=\mathrm{b}_{j k\left(i\right)}$. Independence here leaves cross-year dependence to the copula.

    Previous analyses treated the surveys separately. \citet[][Table S4]{raath2023long} ran snapshot beta regressions for 2003 and 2016 and also modeled the 2016 proportion with 2003 covariates. \citet{burger2023robust} likewise predicted the 2016 value from 2003 covariates, assuming conditional independence between years. These approaches ignore sampling uncertainty in the year-to-year correlation. Modeling the pair $\left(Y_{1i},Y_{2i}\right)$ jointly, with plot-specific random effects and a copula linking rectangular-beta margins, preserves the $\left(0,1\right)$ scale, captures residual between-year dependence, and borrows strength across years when estimating covariate effects.

    With $\mu_{1i} = \mathbb{E}\left(Y_{1i}\right)$ and $\mu_{2i} = \mathbb{E}\left(Y_{2i}\right)$, the logit-scale predictors used later in Eq.~\eqref{eq:linpred} are
    \begin{equation}
      \begin{aligned}
        \operatorname{logit}\left(\mu_{1i}\right) &=
           \beta_{10}
         + \beta_{11}\text{AltMid}_{k\left(i\right)}
         + \beta_{12}\text{AspWest}_{k\left(i\right)}
         + \beta_{13}\text{Agr}_{1i}
         + \beta_{14}\text{LArea}_{1i}
         + \beta_{15}\text{CO}_{1i}
         + \mathrm{b}_{1k\left(i\right)}, \\[4pt]
        \operatorname{logit}\left(\mu_{2i}\right) &=
           \beta_{20}
         + \beta_{21}\text{AltMid}_{k\left(i\right)}
         + \beta_{22}\text{AspWest}_{k\left(i\right)}
         + \beta_{23}\text{Agr}_{2i}
         + \beta_{24}\text{LArea}_{2i}
         + \beta_{25}\text{CO}_{2i}
         + \mathrm{b}_{2k\left(i\right)}.
      \end{aligned}
    \end{equation}
    Here, $\text{Agr}_{ji}$, $\text{LArea}_{ji}$, and $\text{CO}_{ji}$ ($j=1,2$) denote the \textit{Agrostis} cover, log-transformed cushion area, and cover of other vascular or non-vascular plants recorded for cushion $i$ in year $j$. The plot-specific random intercepts $\mathrm{b}_{jk\left(i\right)}$ absorb plot-level heterogeneity, whereas the copula captures residual between-year dependence between $Y_{1i}$ and $Y_{2i}$.

\begin{figure}[!ht]
      \centering
      \includegraphics[width=0.7\linewidth]{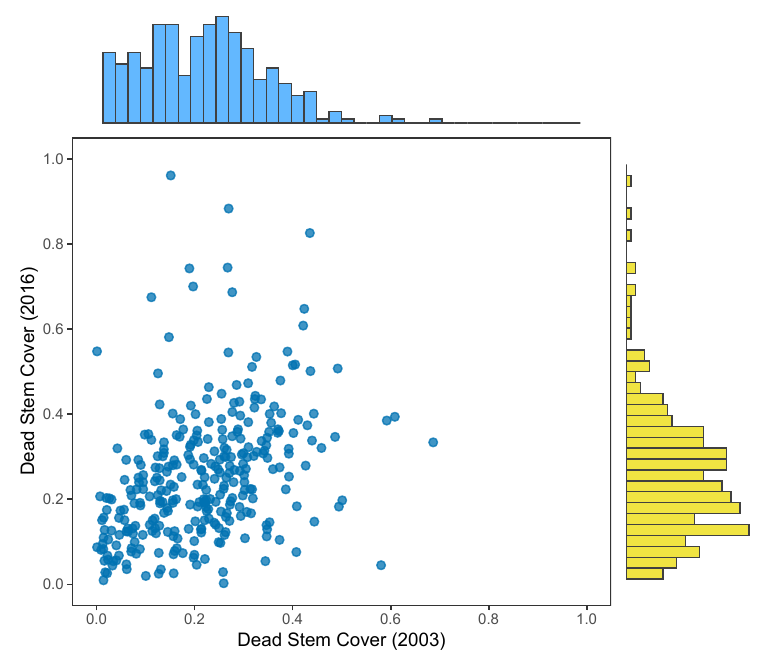}
      \caption{Dead stem cover in 2003 ($Y_{1}$) vs. 2016 ($Y_{2}$) for the 308 cushions. The marginal histograms show a heavier upper tail in 2016, and the scatter indicates a moderate positive concordance, motivating the robust margins and copula link.}
      \label{fig:ds-scatter-overall}
    \end{figure}

    \autoref{fig:ds-scatter-overall} plots the paired responses. A heavier upper tail in 2016 and a clear positive concordance motivate both the rectangular-beta margins and the copula link. \textcolor{red}{Web} \autoref{fig:ds-scatter-byplot} in the Supplementary Information (\textcolor{red}{Web} \autoref{sec:FIG_SUPP}) splits the same scatter by plot, highlighting systematic level shifts that justify the plot-level random effects.

    \section{Bivariate rectangular-beta margins with copula dependence} \label{sec:rectbetacop}
    
    In this section, we define the joint model: each proportion follows a rectangular-beta margin, and the two margins are linked by a one-parameter copula (Gaussian, Gumbel, or Clayton), parameterized by Kendall's $\tau$.
    
    \subsection{Rectangular-beta distribution} \label{sec:rectbeta}
    
    For $a,b>0$, the classical beta density is
    \begin{equation}
      \operatorname{Beta}\left(y; a, b\right) =
      \frac{\Gamma\left(a + b\right)}{\Gamma\left(a\right)\Gamma\left(b\right)}
      y^{a - 1}\left(1 - y\right)^{b - 1}, \quad 0 < y < 1,
    \end{equation}
    with mean $a/\left(a+b\right)$ and precision $a+b$.
    
    Following \citet{Bayes2012}, we say that $Y \sim \mathrm{r\mbox{-}Beta}\left(\mu, \phi, \rho\right)$ if
    \begin{equation}
      f_Y\left(y\left| \mu, \phi, \rho\right.\right) =
      \omega\left(\mu, \phi\right)\mathbf 1_{\left(0,1\right)}\left(y\right)
      + \left\{1 - \omega\left(\mu, \phi\right)\right\}
        \operatorname{Beta}\left(y; \kappa_1\left(\mu, \phi, \rho\right), \kappa_2\left(\mu, \phi, \rho\right)\right),
    \end{equation}
    where $\mathbf 1_{\left(0,1\right)}\left(y\right)=1$ if $0<y<1$ and $0$ otherwise,
    \begin{equation}
      \omega\left(\mu, \phi\right) = \phi\left(1 - \left|2\mu - 1\right|\right), \quad
      \kappa_1\left(\mu, \phi, \rho\right) = \rho \delta\left(\mu, \phi\right), \quad
      \kappa_2\left(\mu, \phi, \rho\right) = \rho\left[1 - \delta\left(\mu, \phi\right)\right],
    \end{equation}
    and
    \begin{equation}
      \delta\left(\mu, \phi\right) =
      \frac{\mu - \tfrac{1}{2} \phi\left(1 - \left|2\mu - 1\right|\right)}
           {1 - \phi\left(1 - \left|2\mu - 1\right|\right)}.
    \end{equation}
    By construction, $0<\omega\left(\mu,\phi\right)<1$ and $0<\delta\left(\mu,\phi\right)<1$ for $\phi\in\left(0,1\right)$. The margin has mean $\mu$ because
    \begin{equation}
      \mathbb{E}\left(Y \left| \mu,\phi,\rho \right.\right)
      = \omega\left(\mu,\phi\right)\cdot \tfrac{1}{2}
        + \left\{1-\omega\left(\mu,\phi\right)\right\}\delta\left(\mu,\phi\right)
      = \mu.
    \end{equation}
    The parameter $\rho>0$ controls precision: a larger $\rho$ reduces dispersion. The weight $\phi\in\left(0,1\right)$ controls $\omega\left(\mu,\phi\right)$, the share of mass shifted from the beta core to a uniform component, thereby flattening the density and down-weighting interior outliers. As $\phi\to0$, the model reduces to the mean-precision beta, $\operatorname{Beta}\left(\mu,\rho\right)\equiv \operatorname{Beta}\left(\mu\rho,\left(1-\mu\right)\rho\right)$.
    
    As $\phi\uparrow1$, the uniform share approaches $1-\left|2\mu-1\right|$ and the density becomes increasingly flat; it is exactly uniform at $\mu=\tfrac12$ and $\phi=1$.
    
    The corresponding cumulative distribution function (CDF) is a convex combination of a uniform and a beta CDF:
    \begin{equation}
      F_Y\left(y\left|\mu,\phi,\rho\right.\right)
      = \omega\left(\mu,\phi\right)y
        + \left\{1-\omega\left(\mu,\phi\right)\right\}
          \operatorname{B}\left(y;\kappa_{1}\left(\mu,\phi,\rho\right), \kappa_{2}\left(\mu,\phi,\rho\right)\right),
    \end{equation}
    where $\operatorname{B}\left(y;a,b\right)\equiv I_{y}\left(a,b\right)$ is the regularized incomplete beta function.
    
    We next couple two such margins through a flexible one-parameter copula, parameterized by Kendall's $\tau$.
    
    \subsection{Copula coupling} \label{sec:copula}
    
    Copulas separate marginal behavior from joint dependence. By Sklar's theorem \citep{VanVliet2023Sklar}, any joint CDF of $\left(Y_{1},Y_{2}\right)\in\left(0,1\right)^{2}$ can be written as
    \begin{equation}
      F_{Y_{1},Y_{2}}\left(y_{1},y_{2}\right)
      = C_{\theta}\left(F_{1}\left(y_{1}\right), F_{2}\left(y_{2}\right)\right),
    \end{equation}
    where $F_{j}$ are the marginal CDFs and $C_{\theta}$ is a bivariate copula with parameter $\theta$. Throughout, $F_{j}\left(\cdot\right)=F_{\mathrm{r\mbox{-}Beta}}\left(\cdot\left|\mu_{j},\phi_{j},\rho_{j}\right.\right)$ as defined in \autoref{sec:rectbeta}. For background, see \citet{Nelsen2006IntroCopulas, Joe2014Dependence}.
    
    Rather than reporting $\theta$ itself, we parameterize each copula by Kendall's rank correlation
    \begin{equation}
      \tau
      =
      \Pr\left[\left(Y_{1}-\tilde{Y}_{1}\right)\left(Y_{2}-\tilde{Y}_{2}\right)>0\right]
      - \Pr\left[\left(Y_{1}-\tilde{Y}_{1}\right)\left(Y_{2}-\tilde{Y}_{2}\right)<0\right],
    \end{equation}
    where $\left(\tilde{Y}_{1},\tilde{Y}_{2}\right)$ is an independent copy of $\left(Y_{1},Y_{2}\right)$. Because $\tau\in\left(-1,1\right)$, it is comparable across families. Tail behavior is summarized by the tail-dependence coefficients of $C_{\theta}$,
    \begin{equation}
      \lambda_{L} = \lim_{q\to0^{+}} \Pr\left(U \le q \left| V \le q\right.\right), \qquad
      \lambda_{U} = \lim_{q\to1^{-}} \Pr\left(U > q \left| V > q\right.\right),
    \end{equation}
    where $U = F_{1}\left(Y_{1}\right)$ and $V = F_{2}\left(Y_{2}\right)$.
    
    We employ three one-parameter copulas: Gaussian (symmetric, $\tau\in\left(-1,1\right)$, no tail dependence), Gumbel (upper-tail dependence, $\tau\in\left(0,1\right)$), and Clayton (lower-tail dependence, $\tau\in\left(0,1\right)$). Within these non-rotated families, negative concordance $\left(\tau<0\right)$ is available only with the Gaussian copula.
    
    Each family admits a closed-form map $\tau \mapsto \theta$ and reduces to independence at $\tau=0$ (Gaussian: $\theta=0$; Gumbel: $\theta=1$; Clayton: $\theta\to0^{+}$). At the opposite extreme, $\theta\to\infty$ in Gumbel and Clayton yields perfect upper- or lower-tail dependence $\left(\lambda_{U}=1\ \text{or}\ \lambda_{L}=1\right)$, respectively.

    Analytic copula expressions, admissible parameter ranges, and the coefficients $\left(\lambda_{L}, \lambda_{U}\right)$ are summarized in \textcolor{red}{Web} \autoref{tab:C-theta-long} of the Supplementary Information (\textcolor{red}{Web} \autoref{sec:TAB_SUPP}).

\begin{figure}[!t]
          \centering
          \subfloat[$\phi_{1}=0,\phi_{2}=0.45$]{
            \begin{minipage}[c]{0.96\textwidth}
              \centering
              \includegraphics[width=0.49\linewidth]
                {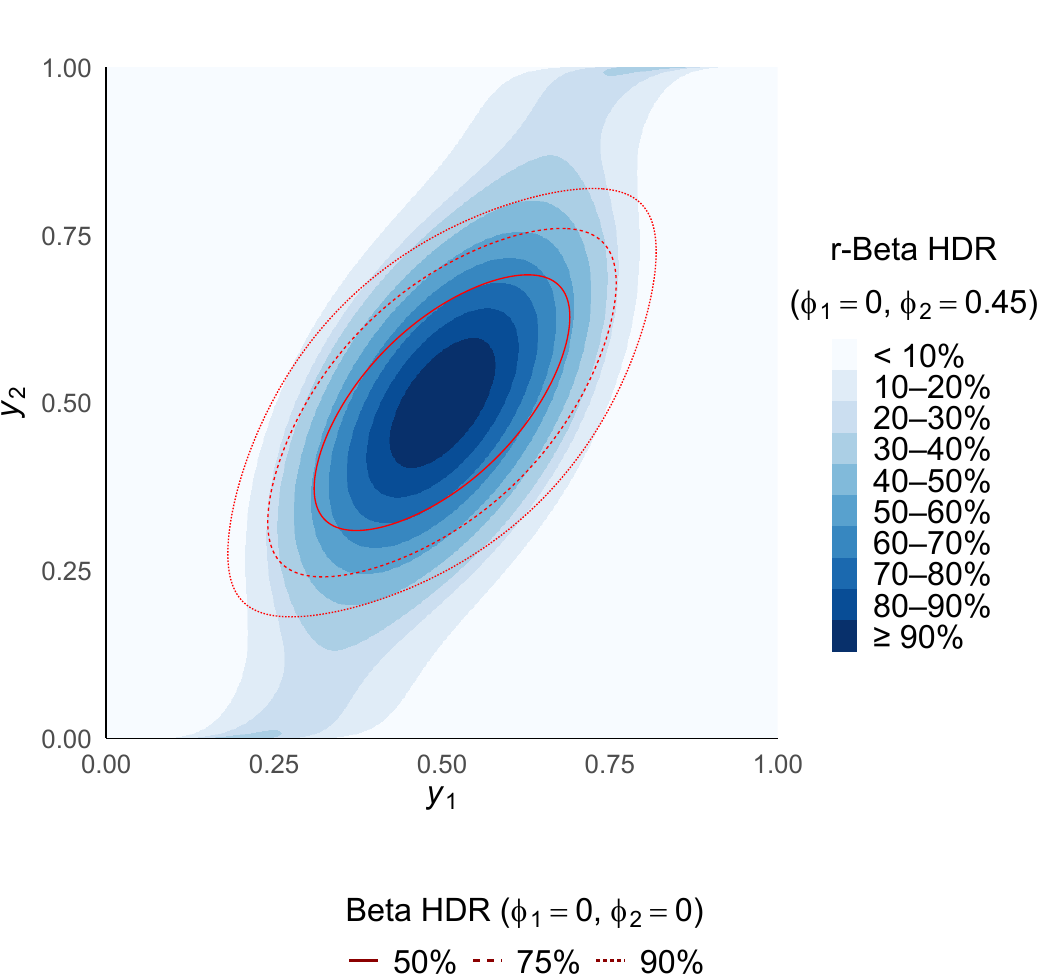}
              \hfill
              \includegraphics[width=0.49\linewidth]
                {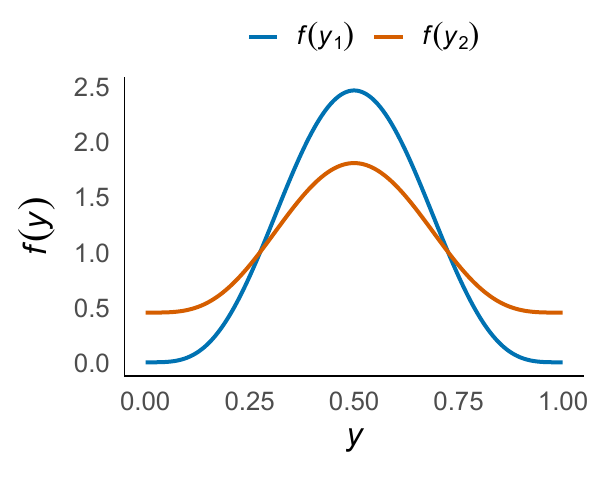}
            \end{minipage}
        }\par\medskip
          \subfloat[$\phi_{1}=0.45,\phi_{2}=0.40$]{
            \begin{minipage}[c]{0.96\textwidth}
              \centering
              \includegraphics[width=0.49\linewidth]
                {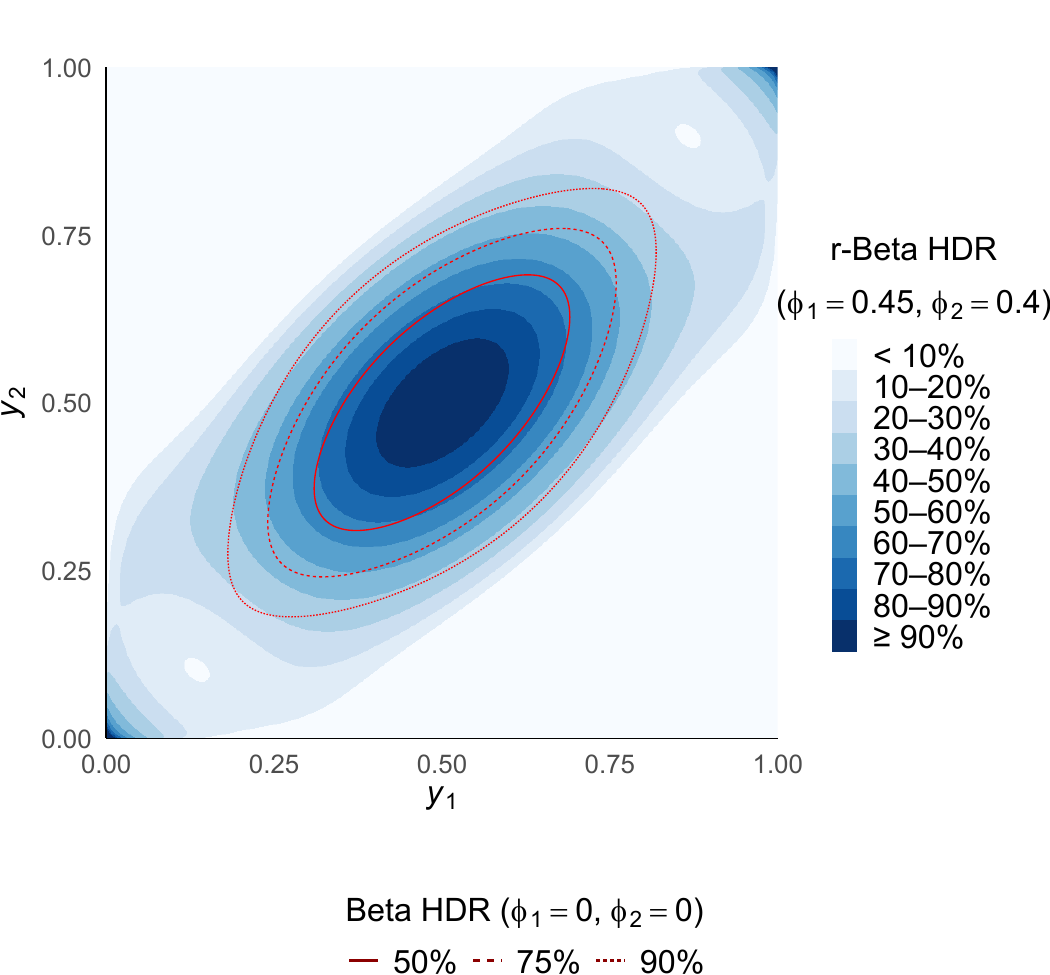}
              \hfill
              \includegraphics[width=0.49\linewidth]
                {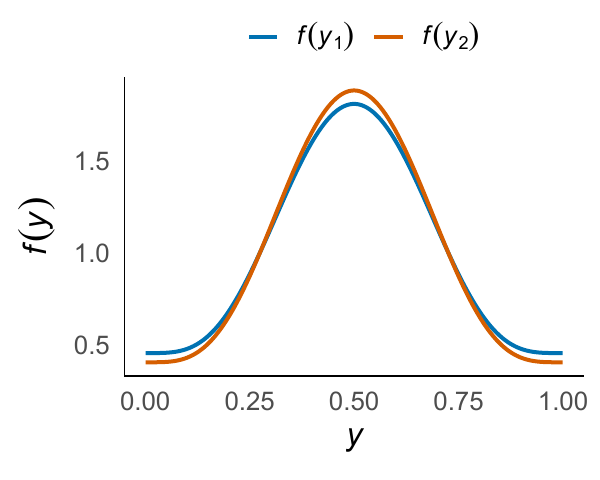}
            \end{minipage}
        }
        \caption{Illustrative effect of the rectangular-beta weights on the joint density $f_{Y_1,Y_2}\left(y_1, y_2\right)$ of Eq.~\eqref{eq:joint-density}. Each row is constructed from a Gaussian copula with Kendall's $\tau=0.40$ and rectangular-beta margins with $\mu_1=\mu_2=0.5$ and precision $\rho=10$. In every left-hand panel, the red curves mark the 50\%, 75\%, and 90\% HDRs for the conventional beta baseline $\left(\phi_1,\phi_2\right)=\left(0,0\right)$, while the blue shading shows HDR bands for the displayed $\left(\phi_1,\phi_2\right)$ values. The right-hand panel in each row gives the corresponding marginal PDFs $f_{Y_1}\left(y_1\right)$ (blue) and $f_{Y_2}\left(y_2\right)$ (orange). Increasing either $\phi_1$ or $\phi_2$ flattens its margin, replacing some of the peaked beta core with a uniform component and thereby producing heavier tails. Rows correspond to $\left(\phi_1,\phi_2\right)=\left(0,0.45\right)$ (top) and $\left(0.45,0.4\right)$ (bottom); the copula dependence is identical across rows, only the marginal flattening differs.} \label{fig:recbeta-gauss-paired}
    \end{figure}

    \autoref{fig:recbeta-gauss-paired} compares $\left(\phi_{1},\phi_{2}\right)=\left(0,0.45\right)$ and $\left(0.45,0.40\right)$ under a Gaussian copula with Kendall's $\tau=0.40$ and margins $\mu_{1}=\mu_{2}=0.5$, $\rho=10$; shaded bands show highest density regions (HDRs) at cumulative probability levels $0.10, 0.20, \ldots, 0.90$ for the rectangular-beta model, and red contours overlay the 50\%, 75\%, and 90\% HDRs for the beta baseline $\left(\phi_{1},\phi_{2}\right)=\left(0,0\right)$.

    Starting from Sklar's identity and assuming absolute continuity,
    \begin{equation}
      F_{Y_{1},Y_{2}}\left(y_{1},y_{2}\right)
      = C_{\theta}\left(F_{Y_{1}}\left(y_{1}\right),F_{Y_{2}}\left(y_{2}\right)\right),
    \end{equation}
    differentiation yields the joint density
    \begin{equation}\label{eq:joint-density}
      f_{Y_{1},Y_{2}}\left(y_{1},y_{2}\right)
      = c_{\theta}\left(F_{Y_{1}}\left(y_{1}\right),F_{Y_{2}}\left(y_{2}\right)\right)
        f_{Y_{1}}\left(y_{1}\right) f_{Y_{2}}\left(y_{2}\right), \qquad 0<y_{1},y_{2}<1,
    \end{equation}
    \newpage
    \clearpage
    where
    \begin{equation}
      c_{\theta}\left(u,v\right) = \frac{\partial^{2}}{\partial u \partial v} C_{\theta}\left(u,v\right)
    \end{equation}
    is the copula density. Let $\bm{\psi}_{j} = \left(\mu_{j},\phi_{j},\rho_{j}\right)$ for $j\in\left\{1,2\right\}$ collect each margin's parameters; then
    \begin{equation}\label{eq:joint-density-psi}
      f_{Y_{1},Y_{2}}\left(y_{1},y_{2}\left| \bm{\psi}_{1},\bm{\psi}_{2},\theta\right.\right)
      = c_{\theta}\left(F_{Y_{1}}\left(y_{1}\left|\bm{\psi}_{1}\right.\right),
                         F_{Y_{2}}\left(y_{2}\left|\bm{\psi}_{2}\right.\right)\right)
        f_{Y_{1}}\left(y_{1}\left|\bm{\psi}_{1}\right.\right)
        f_{Y_{2}}\left(y_{2}\left|\bm{\psi}_{2}\right.\right).
    \end{equation}
    
    \section{Copula mixed-effects regression model} \label{sec:mixedmodel}
    
    Let $i=1,\dots,n$ index units and $j\in\{1,2\}$ the two proportion outcomes. For unit $i$, let the outcome-specific random effects be $\mathbf b_{ji}\in\mathbb R^{q_j}$ and stack
    \begin{equation}
      \mathbf b_i=\left(\mathbf b_{1i}^\top,\mathbf b_{2i}^\top\right)^\top,\qquad
    \mathbf b_i \stackrel{\text{i.i.d.}}{\sim} \mathcal N_{q_1+q_2}\left(\mathbf 0,\mathbf D\right),\qquad
    \mathbf D=\operatorname{diag}\left(\mathbf D_1,\mathbf D_2\right),
    \end{equation}
    where $\mathbf D_j \in \mathbb R^{q_j\times q_j}$ are covariance matrices. We keep $\mathbf D$ block-diagonal so $\operatorname{Cov}\left(\mathbf b_{1i},\mathbf b_{2i}\right)=\mathbf 0$; any residual cross-outcome dependence is handled by the copula. This simplifies interpretation, avoids identifiability issues when both a copula and a cross-block $\mathbf D_{12}$ try to explain the same correlation, and is computationally stable. If a shared latent effect is scientifically warranted, an unstructured $\mathbf D$ can be used.
    
    Conditional on $\mathbf b_i$ and parameters, the joint density follows \autoref{sec:copula} with the margin-specific means
    \begin{equation}\label{eq:linpred}
      \mu_{ji} = \operatorname{logit}^{-1}\left(\mathbf x_{ji}^\top \bm\beta_j + \mathbf z_{ji}^\top \mathbf b_{ji}\right), \qquad j=1,2,
    \end{equation}
    where $\mathbf x_{ji}\in\mathbb R^{p_j}$ collects the fixed-effect covariates (including an intercept), $\bm\beta_j\in\mathbb R^{p_j}$ are the fixed-effect coefficients, and $\mathbf z_{ji}\in\mathbb R^{q_j}$ is the random-effect design vector, so that
    \begin{equation}\label{eq:mixed-joint}
      f_{Y_{1},Y_{2}}\left(y_{1i}, y_{2i} \left| \mathbf b_i, \bm\psi_{1i}, \bm\psi_{2i}, \theta \right.\right)
      = c_\theta\left(
          F_{Y_1}\left(y_{1i} \left| \bm\psi_{1i} \right.\right),
          F_{Y_2}\left(y_{2i} \left| \bm\psi_{2i} \right.\right)
        \right)
        \prod_{j=1}^{2} f_{Y_j}\left(y_{ji} \left| \bm\psi_{ji} \right.\right).
    \end{equation}
    Here $f_{Y_j}$ and $F_{Y_j}$ are the rectangular-beta density and CDF from \autoref{sec:rectbeta}, $c_\theta$ is the copula density, and $\bm\psi_{ji}=\left(\mu_{ji},\phi_j,\rho_j\right)$ with $\mu_{ji}$ defined in Eq.~\eqref{eq:linpred}.

    For margin $j$, the precision $\rho_j>0$ controls dispersion and the weight $\phi_j\in\left(0,1\right)$ down-weights interior observations in the rectangular-beta margin; dependence is indexed by the copula parameter $\theta$ (equivalently Kendall's $\tau$). In the baseline model, $\phi_j$ and $\rho_j$ are constant across units; if needed, they can be linked to covariates.
    
    Setting $\phi_1=\phi_2=0$ recovers a beta-margin copula mixed model \citep{Joe2014Dependence}. Independence occurs at $\tau=0$ ($\theta=0$ for Gaussian, $\theta=1$ for Gumbel, and $\theta\to0^{+}$ for Clayton). We work on the $\tau$ scale throughout and map to $\theta$ using \textcolor{red}{Web} \autoref{tab:C-theta-long} (\textcolor{red}{Web} \autoref{sec:TAB_SUPP}).
    
    We compare five specifications: $\text{Beta}_{\text{Indep}}$ (beta margins, independent), $\text{RectBeta}_{\text{Indep}}$ (rectangular-beta margins, independent), $\text{RectBeta}_{\text{Gauss}}$, $\text{RectBeta}_{\text{Gumbel}}$, $\text{RectBeta}_{\text{Clayton}}$. All share the same fixed- and random-effects structure in Eq.~\eqref{eq:linpred}.

    \section{Inference} \label{sec:inference}

    We work in a fully Bayesian framework, combining the likelihood of the mixed-effects copula model with weakly informative, mutually independent priors.

    For each margin, the fixed-effect vector is given a weakly informative normal prior  
    \begin{equation}
        \bm{\beta}_{j} \sim \mathcal{N}_{p_{j}} \left(\mathbf{0}, 100^{2} \mathbf{I}_{p_{j}}\right), \quad j = 1, 2,
    \end{equation}
    so every regression coefficient has prior mean $0$ and standard deviation $100$.

    If a margin contains only one random effect ($q_{j} = 1$), the associated standard deviation receives a half-$t$ prior with location $0$, scale $2.5$, and $3$ degrees of freedom,
    \begin{equation} \label{eq:half-t}
        \sigma_{j} \sim t^{+}_{3} \left(0, 2.5\right), \quad j = 1, 2.
    \end{equation}
    (This choice places 95\% of the prior mass below $\sigma_{j}\approx 7.95$ on the logit scale, so it is effectively non-informative over the unit interval.)
    
    For higher-dimensional random effects, one may decompose the covariance matrix as  
    \begin{equation}
        \mathbf{D}_{j} = \operatorname{diag} \left(\bm{\sigma}_{j}\right) \mathbf{R}_{j} \operatorname{diag} \left(\bm{\sigma}_{j}\right),
    \end{equation}
    assign the same half-$t$ prior in Eq.~\eqref{eq:half-t} to every element of $\bm{\sigma}_{j}$, and use an $\mathrm{LKJ}$ prior for the correlation matrix $\mathbf{R}_{j}$ \citep{lewandowski2009generating}. (The data example that follows involves only a single random effect, so $\mathbf{R}_{j}$ drops out.)

    Rectangular-beta weights and precisions are given standard diffuse choices:
    \begin{equation}
        \phi_{j} \sim \mathcal{U} \left(0, 1\right), \quad
        \rho_{j} \sim \operatorname{Gamma} \left(0.0001, 0.0001\right), \quad j = 1, 2,
    \end{equation}
    where $\operatorname{Gamma} \left(a, b\right)$ is parameterized by shape $a$ and rate $b$. If weaker mixing arises, the diffuse priors may instead be set to mildly informative ones to reduce the potential trade-off between $\rho_{j}$ and $\phi_{j}$.

    Dependence is described through Kendall's rank correlation $\tau$. A flat prior is used on its feasible interval,
    \begin{equation}
        \tau \sim
        \begin{cases}
            \mathcal{U} \left(-1, 1\right), & \text{Gaussian copula}, \\
            \mathcal{U} \left(0, 1\right), & \text{Gumbel or Clayton copula},
        \end{cases}
    \end{equation}
    and the copula parameter $\theta$ is obtained deterministically from $\tau$ via the family-specific mappings in \textcolor{red}{Web} \autoref{tab:C-theta-long} (\textcolor{red}{Web} \autoref{sec:TAB_SUPP}). Working on the $\tau$-scale makes results comparable across copula families.

    With data $\mathcal{Y} = \left\{\left(y_{1i}, y_{2i}\right)\right\}_{i = 1}^{n}$ and random effects $\mathbf{b}_{i}$, the complete data log-likelihood is
    \begin{equation}\label{eq:loglik}
        \ell(\mathcal{Y};\bm\Theta) =
        \sum_{i = 1}^{n} \log \left[
        f_{Y_{1}, Y_{2}} \left(y_{1i}, y_{2i}\left|
        \bm{\psi}_{1i}, \bm{\psi}_{2i}, \theta\left(\tau\right)\right.\right)
        \right]
        + \sum_{i = 1}^{n} \log \left[
        \varphi_{q_{1} + q_{2}} \left(\mathbf{b}_{i}; \mathbf{0}, \mathbf{D}\right)
        \right],
    \end{equation}
    where $\bm\Theta=\left(\bm\beta_{1},\bm\beta_{2},\phi_{1},\phi_{2},\rho_{1},\rho_{2},\tau,\mathbf D,\mathbf b_{1:n}\right)$ and $\varphi_{d}\left(\cdot; \mathbf{0}, \mathbf{D}\right)$ denotes the density of $\mathcal{N}_{d} \left(\mathbf{0}, \mathbf{D}\right)$. The posterior density is proportional to the product of this likelihood and the prior densities, which factorize as
    \begin{equation}
        \pi \left(\bm{\Theta}\right) =
        \pi \left(\bm{\beta}_{1}\right)
        \pi \left(\bm{\beta}_{2}\right)
        \pi \left(\phi_{1}\right)
        \pi \left(\phi_{2}\right)
        \pi \left(\rho_{1}\right)
        \pi \left(\rho_{2}\right)
        \pi \left(\tau\right)
        \pi \left(\mathbf{D}\right).
    \end{equation}
    Posterior sampling can be carried out with generic Markov chain Monte Carlo (MCMC) software such as \texttt{JAGS} \citep{PLUMMER2003}. All numerical results reported later were obtained with \texttt{JAGS}.

    \section{Model comparison} \label{sec:modelcomp}

    Model complexity and fit are weighed using the LML (log-evidence) of a candidate model $\mathcal{M}$:
    \begin{equation}
        \log \left\{p\left(\mathcal{Y}\left|\mathcal{M}\right.\right)\right\} =
        \log \left(\int
        \exp \left\{\ell \left(\mathcal{Y}; \bm{\Theta}\right)\right\}
        \pi \left(\bm{\Theta}\right) \mathrm{d} \bm{\Theta}\right),
    \end{equation}
    where the complete-data log-likelihood $\ell \left(\mathcal{Y}; \bm{\Theta}\right)$ is given in Eq.~\eqref{eq:loglik}. Because the integral averages the likelihood under the prior, the LML integrates over all random effects and hyperparameters; it rewards fit while (prior-dependently) penalizing unnecessary flexibility. Between any two candidates, the model with the larger LML is preferred; a difference of two natural-log units corresponds to a Bayes factor of $\mathrm e^{2} \approx 7.4$, often viewed as ``positive" evidence \citep{Kass1995}.

    We estimate the integral via bridge sampling using the \texttt{bridgesampling} package \citep{gronau2020computing, meng1996simulating}. In practice, we provide (i) the joint log-likelihood $\ell \left(\mathcal{Y}; \bm{\Theta}\right)$ and (ii) the log-prior $\log\left\{\pi\left(\bm{\Theta}\right)\right\}$; the method combines these with MCMC draws and returns the log-evidence.

    \section{Model diagnostics} \label{sec:diag}

    Model adequacy is examined from two complementary angles: (i)~marginal fit of each outcome, assessed with simulation-based residuals, and (ii)~adequacy of the dependence between the two outcomes, assessed with empirical tail and quadrant functions.

    \subsection{Posterior predictive residual checks} \label{sec:pp_residuals}

    Model adequacy is assessed with the simulation-based (rank) residuals implemented in \texttt{DHARMa} \citep{HARTIG2021A}, a generalization of the randomized quantile residuals of \citet{DUNN1996} to mixed-effects and Bayesian settings.
    
    For each retained MCMC draw $\bm{\Theta}^{\left(s\right)}$, $s=1,\ldots,S$, we generate a full posterior predictive replicate $\tilde{Y}_{1i}^{\left(s\right)}, \tilde{Y}_{2i}^{\left(s\right)}$. Collecting the simulations in an $n\times S$ matrix, \texttt{DHARMa} assigns to every observation a scaled rank residual:
    \begin{equation}
        R_{ji} =
        \frac{\displaystyle\sum_{s = 1}^{S} \mathbf{1} \left\{\tilde{Y}_{ji}^{\left(s\right)} < Y_{ji}\right\} + u_{ji}}{S + 1}, \quad
        u_{ji} \sim \mathcal{U} \left(0, 1\right), \quad
        j = 1, 2, \quad i = 1, \ldots, n.
    \end{equation}
    Under the fitted model, the residuals $R_{ji}$ are i.i.d. $\mathcal{U}\left(0, 1\right)$.

    We apply three formal residual tests to each margin (via \texttt{DHARMa}): (i) a Kolmogorov-Smirnov-type test of the uniformity of $R_{ji}$; (ii) a dispersion test comparing the variance of $R_{ji}$ with that expected under the fitted model; and (iii) an outlier test based on the tails of the simulated residual distribution.
    
    \subsection{Empirical validation of the copula dependence} \label{sec:gof_copula}

    To gauge how well a candidate copula captures dependence, we analyze functions of the bivariate probability integral transform (PIT) pairs $\left(U_{1i},U_{2i}\right)$, where each coordinate is marginally $\mathcal{U}\left(0,1\right)$ under the fitted model. For upper-tail association (e.g., Gumbel), we monitor
    \begin{equation}
      \chi\left(u\right) = \Pr\left(U_{1} > 1 - u \left| U_{2} > 1 - u \right.\right), \quad 0 < u < 1,
    \end{equation}
    and for lower-tail behavior (e.g., Clayton), the analogue
    \begin{equation}
      \chi_{L}\left(u\right) = \Pr\left(U_{1} \le u \left| U_{2} \le u \right.\right), \quad 0 < u < 1.
    \end{equation}
    Overall concordance is summarized by the quadrant probability
    \begin{equation}
      K\left(u\right) = \Pr\left(U_{1} \le u, U_{2} \le u\right).
    \end{equation}
    For each retained MCMC draw $\bm{\Theta}^{\left(s\right)}$, $s = 1,\ldots,S$, we compute PITs $U_{ji}^{\left(s\right)} = F_{Y_j}\left(y_{ji} \left| \bm{\psi}_{ji}^{\left(s\right)} \right.\right)$ using the corresponding random-effects draws, then form empirical curves $\widehat{\chi}^{\left(s\right)}\left(u\right)$, $\widehat{\chi}_{L}^{\left(s\right)}\left(u\right)$, and $\widehat{K}^{\left(s\right)}\left(u\right)$ on a grid $u \in \{0.05, 0.10, \ldots, 0.95\}$. We report the posterior means $\bar{\chi}\left(u\right) = S^{-1} \sum_{s=1}^{S} \widehat{\chi}^{\left(s\right)}\left(u\right)$ (and analogues for $\chi_{L}$ and $K$).

    Pointwise posterior predictive envelopes are built as follows. For each replicate $b = 1,\ldots,B$, (i) draw the copula parameter from its posterior, (ii) simulate $n$ bivariate PIT pairs from the implied copula, and (iii) recompute $\widehat{\chi}^{\left(b\right)}\left(u\right)$ (or $\widehat{\chi}_{L}^{\left(b\right)}\left(u\right)$) and $\widehat{K}^{\left(b\right)}\left(u\right)$. At every $u$, the 2.5\% and 97.5\% quantiles of the $B$ curves define the envelope.

    If the empirical curves remain within their envelopes across the range of $u$, the copula is judged adequate; systematic departures indicate a lack of fit.

    \section{Application} \label{sec:application}

    We fitted the copula mixed models introduced in \autoref{sec:mixedmodel} under the Bayesian framework described in \autoref{sec:inference} to the cushion plant dataset in \autoref{sec:dataset}. Implementation is summarized below.

    \subsection{Computational details}

    All models were run in \texttt{R} with \texttt{JAGS} using four independent chains of 40~000 iterations. The first 15~000 iterations were discarded as burn-in; the remainder were thinned every 25\textsuperscript{th} draw, giving 1~000 posterior samples per chain and 4~000 usable draws in total. Convergence was satisfactory, with potential scale reduction factor values not exceeding 1.05.

    For $j = 1,2$ and cushion $i$, let
    \begin{equation}
      \mathbf{x}_{ji}^{\top} =
      \left(
        1, 
        \text{AltMid}_{k\left(i\right)}, 
        \text{AspWest}_{k\left(i\right)}, 
        \text{Agr}_{ji}, 
        \text{LArea}_{ji}, 
        \text{CO}_{ji}
     \right),
      \qquad
      \bm{\beta}_{j} =
      \left(
        \beta_{j0}, \beta_{j1}, \beta_{j2}, \beta_{j3}, \beta_{j4}, \beta_{j5}
     \right)^{\top}.
    \end{equation}
    In this application, we consider the special case with plot-level random intercepts, where $z_{ji}=1$ and $\mathrm{b}_{ji}=\mathrm{b}_{j k\left(i\right)}$.
    
    Posterior medians are reported with 95\% highest posterior density (HPD) intervals. LMLs were estimated with the \texttt{warp3} bridge sampler in \texttt{bridgesampling}. \texttt{DHARMa} residual checks were applied to each margin. For the copula diagnostics, envelopes for $\chi\left(u\right)$, $\chi_{L}\left(u\right)$, and $K\left(u\right)$ used $B = 500$ posterior draws of the copula parameter, with $n$ simulated PIT pairs per draw.

    All code is available at \url{https://github.com/DABURGER1/r-Beta-Copula}. Computations were run on an 11\textsuperscript{th} generation Intel Core i5-1145G7 processor (2.60 GHz) with 16~GB RAM.

    \subsection{Model ranking and diagnostics}

    \autoref{tab:model-variants-results} lists LMLs and \texttt{DHARMa} residual \emph{p}-values. The copula variants have the largest LMLs, with the Gumbel model highest and the Gaussian model within 0.9 log units; the Clayton model is lower by about 11 log units, and both independence models have markedly smaller support. Every rectangular-beta specification, whether independent or copula-linked, passes the uniformity, dispersion, and outlier tests ($p \ge 0.05$). In contrast, the conventional beta model fails at least one test in each year, highlighting the need for a robust margin. Overall, rectangular-beta margins handle outliers effectively, and adding a copula further improves fit; the Gumbel ranks first by LML, with the Gaussian essentially comparable.

    \setlength{\tabcolsep}{0.07cm}
    \begin{longtable}{l c r r r r r r}
        \caption{LMLs and residual test \emph{p}-values for competing models.} 
        \label{tab:model-variants-results}\\
        \toprule
        & & \multicolumn{6}{c}{{\bf Residual test}} \\
        \cmidrule(lr){3-8}
        \multicolumn{2}{c}{} & \multicolumn{3}{c}{\textbf{\emph{p}-value (2003)}} & \multicolumn{3}{c}{\textbf{\emph{p}-value (2016)}} \\
        \cmidrule(lr){3-5} \cmidrule(lr){6-8}
        \textbf{Model} & \textbf{LML} & \textbf{Uniform} & \textbf{Dispersion} & \textbf{Outliers} & \textbf{Uniform} & \textbf{Dispersion} & \textbf{Outliers} \\
        \midrule
        \endfirsthead
        
        \caption[]{(continued)} \\
        \toprule
        & & \multicolumn{6}{c}{{\bf Residual test}} \\
        \cmidrule(lr){3-8}
        \multicolumn{2}{c}{} & \multicolumn{3}{c}{\textbf{\emph{p}-value (2003)}} & \multicolumn{3}{c}{\textbf{\emph{p}-value (2016)}} \\
        \cmidrule(lr){3-5} \cmidrule(lr){6-8}
        \textbf{Model} & \textbf{LML} & \textbf{Uniform} & \textbf{Dispersion} & \textbf{Outliers} & \textbf{Uniform} & \textbf{Dispersion} & \textbf{Outliers} \\
        \midrule
        \endhead
        
        \midrule
        \endfoot
        
        \bottomrule
        \endlastfoot
        
        $\text{Beta}_{\text{Indep}}$ & 318.451 & 0.041 & 0.122 & 0.143 & 0.048 & 0.533 & $<$0.001 \\ 
          $\text{RectBeta}_{\text{Indep}}$ & 339.763 & 0.160 & 0.147 & 0.143 & 0.945 & 0.638 & 1.000 \\ 
          $\text{RectBeta}_{\text{Gauss}}$ & 361.781 & 0.187 & 0.117 & 0.143 & 0.942 & 0.594 & 1.000 \\ 
          $\text{RectBeta}_{\text{Gumbel}}$ & 362.689 & 0.148 & 0.050 & 1.000 & 0.881 & 0.307 & 1.000 \\ 
          $\text{RectBeta}_{\text{Clayton}}$ & 351.697 & 0.109 & 0.070 & 0.143 & 0.942 & 0.467 & 1.000 \\ 
    \end{longtable}

\begin{figure}[!ht]
      \centering
      \captionsetup[subfigure]{margin=0pt,skip=2pt}
      \begin{subfigure}[b]{0.49\linewidth}
        \centering
        \includegraphics[width=\linewidth]{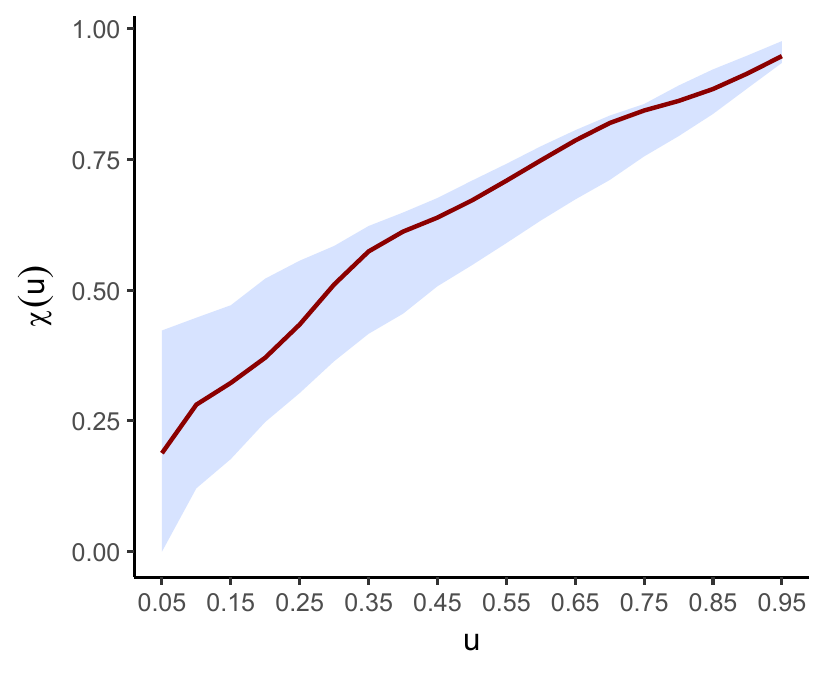}
        \caption{$\chi\left(u\right)$ envelope}
      \end{subfigure}
      \hfill
      \begin{subfigure}[b]{0.49\linewidth}
        \centering
        \includegraphics[width=\linewidth]{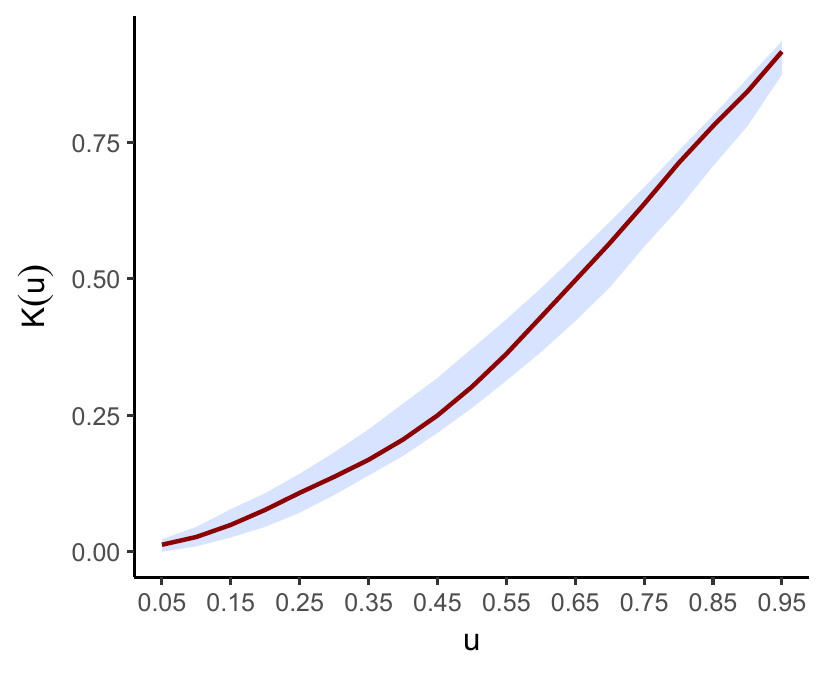}
        \caption{$K\left(u\right)$ envelope}
      \end{subfigure}
      \caption{Goodness-of-fit diagnostics for the Gaussian copula model: The $\chi\left(u\right)$ function (left) probes tail association while the $K\left(u\right)$ function (right) assesses overall quadrant probabilities. Solid red curves represent the empirical statistics, while the blue ribbons show 95\% posterior predictive envelopes. Curves remaining inside the ribbons indicate that the fitted model reproduces both tail behavior and global dependence adequately.}
      \label{fig:Gauss-gof-chiK}
    \end{figure}

    \autoref{fig:Gauss-gof-chiK} compares the empirical $\chi\left(u\right)$ and $K\left(u\right)$ functions with 95\% posterior predictive envelopes for the Gaussian copula. Analogous diagnostics for the Gumbel and Clayton copulas are provided in the Supplementary Information (\textcolor{red}{Web}~\autoref{sec:FIG_SUPP}; see \textcolor{red}{Web}~\autoref{fig:Gumbel-gof-chiK} and \textcolor{red}{Web}~\autoref{fig:Clayton-gof-chiK}). All three models reproduce the observed dependence: the Gaussian and Gumbel curves lie entirely within their envelopes, whereas the Clayton shows a minor outward excursion of $\chi_{L}\left(u\right)$ and $K\left(u\right)$ near $u \approx 0.7$.
    
    We regard every copula as adequate; the Gumbel attains the highest LML, with the Gaussian a close second.

    \subsection{Posterior estimates of covariate effects}

    \autoref{fig:betaHPD-2016} displays posterior medians and 95\% HPD intervals for the six fixed effect coefficients in 2016. The analogous plot for 2003 is available in the Supplementary Information (\textcolor{red}{Web} \autoref{sec:FIG_SUPP}; see \textcolor{red}{Web} \autoref{fig:betaHPD-2003}). \textcolor{red}{Web} \autoref{tab:posterior} in the Supplementary Information (\textcolor{red}{Web} \autoref{sec:TAB_SUPP}) lists the posterior medians and associated 95\% HPD intervals for every model parameter for each of the five fitted specifications.
    
    Across all five models, the 2003 coefficients remain virtually unchanged; point estimates shift only slightly, and their 95\% HPD intervals overlap almost completely. In 2016, however, rectangular-beta margins and the copula reshape the estimates: the robust margin reduces the influence of outliers, and the copula further adjusts coefficients in line with year-to-year correlation. Thus, accounting for residual dependence matters for 2016, while the choice of copula family only refines the broad patterns revealed by the rectangular-beta margin.

    \afterpage{
        \begin{landscape}
        
          \begin{figure}[!ht]
            \centering
            \includegraphics{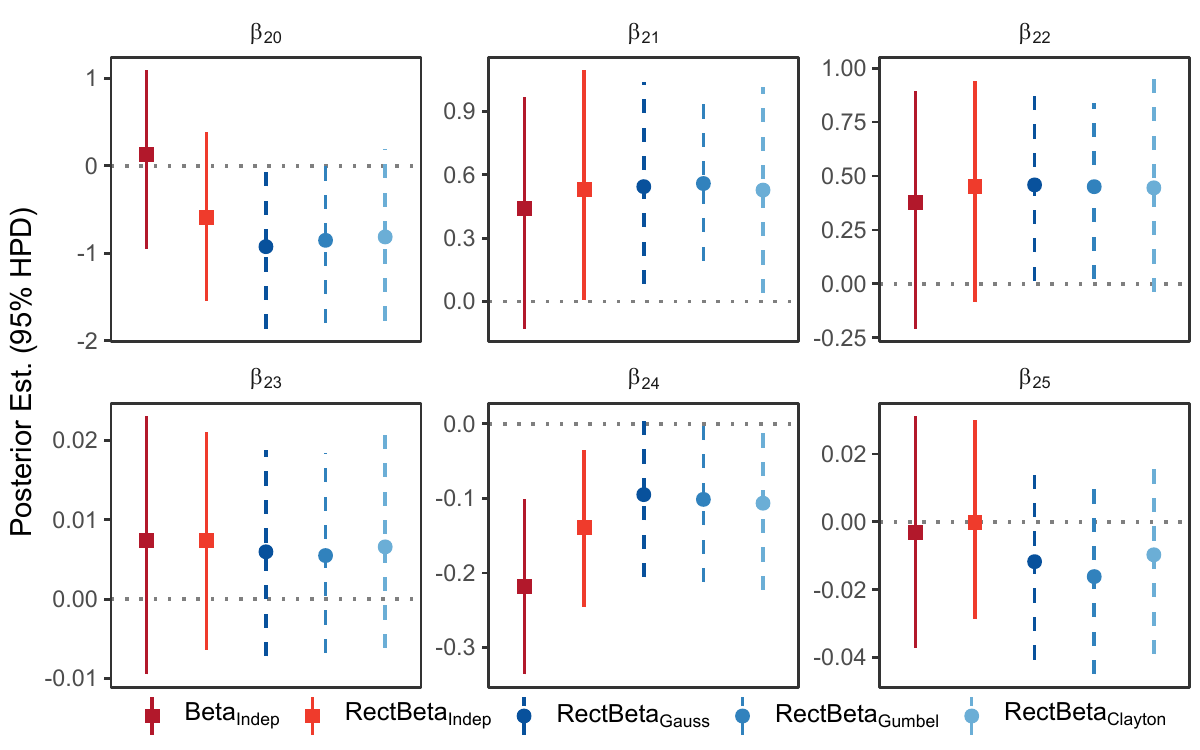}
            \caption{Posterior medians (symbols) and 95\% HPD intervals (vertical bars) for the six fixed-effect coefficients, 2016 outcome. Squares: independent-margin models; circles: copula models. Colors: dark red--light red for conventional vs. rectangular-beta independent models; dark--medium--light blue for Gaussian, Gumbel, and Clayton copulas, respectively. Solid lines denote independent margins, dashed lines denote copulas. The grey dotted horizontal line marks the zero-effect level; intervals that cross this line include zero in their 95\% HPD and, therefore, do not provide evidence of a non-zero effect. Coefficient definitions (design matrix order): $\beta_{20}$:~intercept; $\beta_{21}$:~$\text{AltMid}_{k\left(i\right)}$ -- indicator for mid- vs. high-elevation plot; $\beta_{22}$:~$\text{AspWest}_{k\left(i\right)}$ -- indicator for west-facing vs. east-facing slope; $\beta_{23}$:~$\text{Agr}_{2i}$ -- percent cover of \textit{Agrostis magellanica}; $\beta_{24}$:~$\text{LArea}_{2i}$ -- $\log\left(\text{cushion area}\right)$; $\beta_{25}$:~$\text{CO}_{2i}$ -- percent cover of all other vascular and non-vascular plants.}
            \label{fig:betaHPD-2016}
          \end{figure}
          
        \end{landscape}
    }

    \section{Simulation study} \label{sec:simulation}

    To gauge the frequentist reliability of the Gaussian-rectangular-beta specification, we generated artificial datasets under six data-generating settings that vary the marginal shape parameters $\left(\phi_{1}, \phi_{2}\right) \in \left\{\left(0.05, 0.05\right), \left(0.20, 0.05\right), \left(0.20, 0.20\right) \right\}$ and the copula parameter $\tau \in \left\{0, 0.25\right\}$, and then repeated each setting for two sample sizes ($n = 300$ and $n = 500$). Random effects were omitted to keep the design minimal. The fixed coefficients were set to $\beta_{11} = \operatorname{logit}\left(0.30\right)$, $\beta_{12} = 0.3$, $\beta_{21} = \operatorname{logit}\left(0.60\right)$, $\beta_{22} = -0.3$, with dispersion parameters $\rho_{1} = \rho_{2} = 50$. For every configuration, we produced 200 replicate datasets and analyzed each with the exact MCMC workflow used in the real data application.

    \textcolor{red}{Web} \autoref{tab:bias_rmse_cov} in the Supplementary Information (\textcolor{red}{Web} \autoref{sec:TAB_SUPP}) reports the empirical bias, root mean square error (RMSE), and 95\% HPD coverage. Across all scenarios, the model performs well, with biases negligible relative to sampling variability. The RMSEs decrease as expected when the sample size increases from 300 to 500, and empirical coverages remain close to the nominal 0.95 level.

    \section{Discussion}\label{sec:discussion}

    Rectangular-beta margins combined with a one-parameter copula provide a flexible, parsimonious tool for bounded proportions. In the \textit{Azorella} study, \texttt{DHARMa} diagnostics improved markedly and LMLs increased by more than 20 natural-log units relative to an independence model. Gaussian and Gumbel copulas fitted best, while Clayton ranked lower yet remained adequate by the dependence diagnostics. Simulation results across a grid of tail weight and dependence settings showed negligible bias, near-nominal 95\% HPD coverage, and decreasing RMSE with sample size, suggesting the Gaussian-rectangular-beta specification is a reliable default for paired proportions with occasional interior extremes.

    Accounting for year-to-year dependence alters ecological conclusions. With dependence modeled, the west-slope effect ($\beta_{22}$) in 2016 is credibly positive, whereas independence yields no effect. Conversely, the negative link between log cushion area and dead stem cover ($\beta_{24}$) weakens, with its 95\% HPD interval near zero, illustrating how ignoring residual correlation can mislead inference.

    Future directions include: (i) investigating a $t$ copula to capture symmetric tail clustering, noting that the extra degrees of freedom parameter is weakly identifiable here; (ii) accommodating boundary mass at 0 or 1 via hurdle or zero-one-inflated margins; (iii) extending beyond two responses using vine or factor copulas with shrinkage priors; and (iv) relaxing the independence of plot intercepts across years to an unstructured $2\times2$ block when a shared latent driver is plausible.

    Overall, the framework is straightforward to implement in standard Bayesian software, yields interpretable parameters, and performs well under residual tests and simulation diagnostics (bias and coverage).

    \setlength{\bibhang}{1.2em}
    
    \renewcommand{\refname}{References}
    \bibliographystyle{unsrt}
    \bibliography{Bibliography.bib}
    
    %\section*{Acknowledgments}
    
    %This work is based on the research supported by the National Research Foundation (NRF) of South Africa (Grant number 132383). Opinions expressed and conclusions arrived at are those of the authors and are not necessarily to be attributed to the NRF.

    \section*{Conflict of interest}

    No conflicts of interest have been declared.

    \newpage
    \clearpage
    
    \setcounter{page}{1}
    \setcounter{figure}{0}
    \setcounter{table}{0}
    \renewcommand{\restoreapp}{}
    \renewcommand\appendixname{Web Appendix}
    \renewcommand{\figurename}{WEB FIGURE}
    \renewcommand{\tablename}{WEB TABLE}
    \renewcommand{\headrulewidth}{0pt}
    \titleformat{\section}{\large\bfseries}{\appendixname~\thesection .}{0.5em}{}
    
    \newpage
    \clearpage
    
    \begin{appendices}

        \onehalfspacing
        
        {\Large\bf Outlier-robust copula regression for bivariate continuous proportions: an application to cushion plant vitality}
        
        %\vskip 1.0cm
        
        %{\normalsize Divan A. Burger,
        %Janet van Niekerk,
        %Peter C. le Roux,
        %Morgan J. Raath-Krüger}
        
        \vskip 4.5truecm
        
        \begin{center}
            \noindent
            {\Large\bf Supporting Information}
        \end{center}

        \begin{landscape}

            \section{Figures} \label{sec:FIG_SUPP}

            \begin{figure}[H]
                \centering
                \includegraphics[width=\linewidth]
                {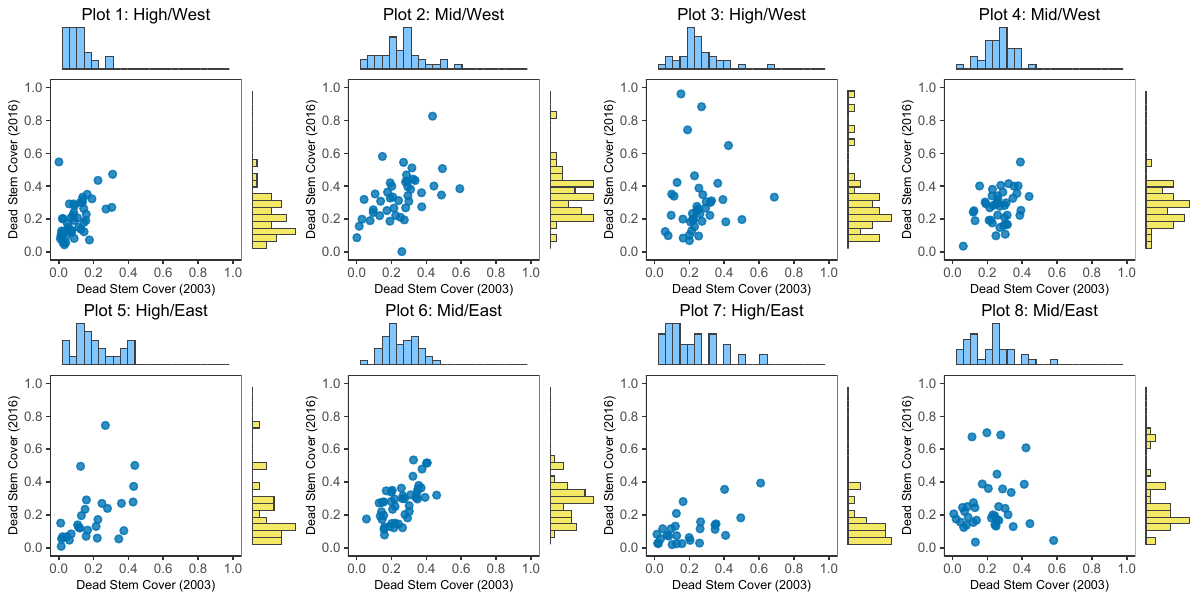}
                \caption{Dead stem cover by plot. Each panel displays the scatter plot of $Y_{1}$ vs. $Y_{2}$ for one plot (the plot number and altitude/aspect combination appear in the strip heading), together with its marginal histograms. Location shifts between panels confirm the need for plot-specific random intercepts, while the broadly similar cloud shapes support a common copula dependence structure.} \label{fig:ds-scatter-byplot}
            \end{figure}
            
        \end{landscape}

        \begin{figure}
          \centering
          \begin{subfigure}[b]{0.49\linewidth}
            \centering
            \includegraphics[width=\linewidth]{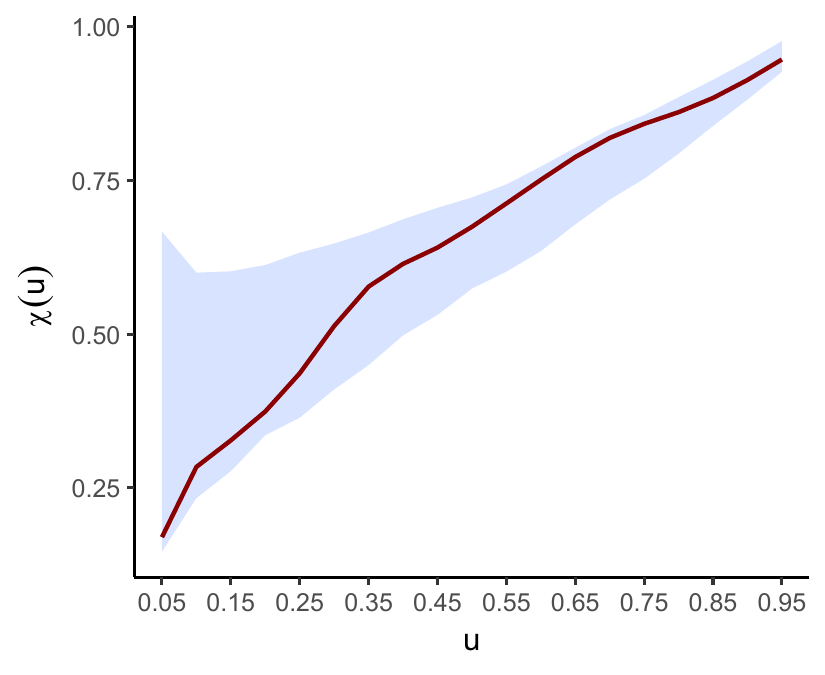}
            \caption{$\chi\left(u\right)$ envelope}
          \end{subfigure}
          \hfill
          \begin{subfigure}[b]{0.49\linewidth}
            \centering
            \includegraphics[width=\linewidth]{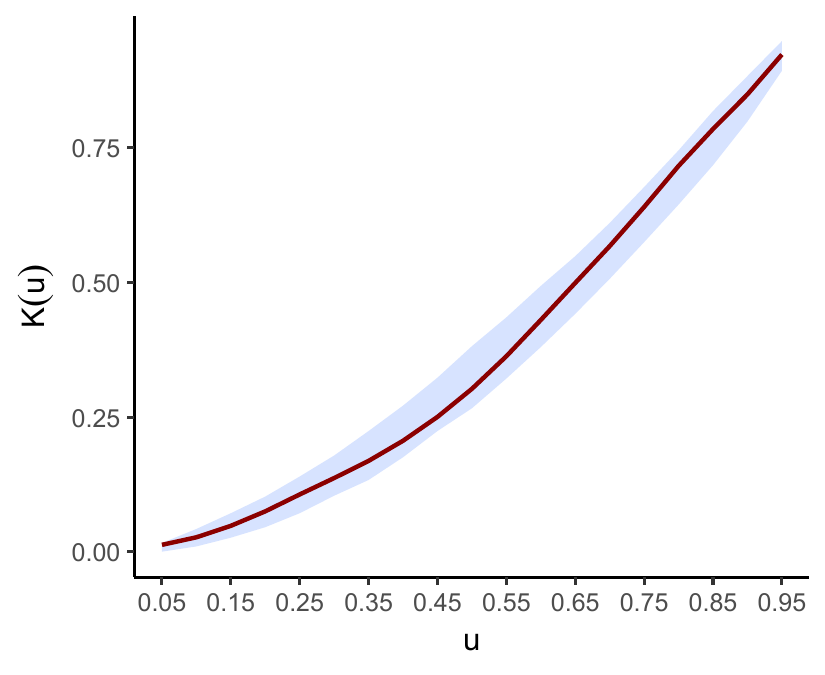}
            \caption{$K\left(u\right)$ envelope}
          \end{subfigure}
          \caption{Goodness-of-fit diagnostics for the Gumbel copula model: The $\chi\left(u\right)$ function (left) probes tail association while the $K\left(u\right)$ function (right) assesses overall quadrant probabilities. Solid red curves represent the empirical statistics, while the blue ribbons show 95\% posterior predictive envelopes. Curves remaining inside the ribbons indicate that the fitted model reproduces both tail behavior and global dependence adequately.}
          \label{fig:Gumbel-gof-chiK}
        \end{figure}
    
        \begin{figure}
          \centering
          \begin{subfigure}[b]{0.49\linewidth}
            \centering
            \includegraphics[width=\linewidth]{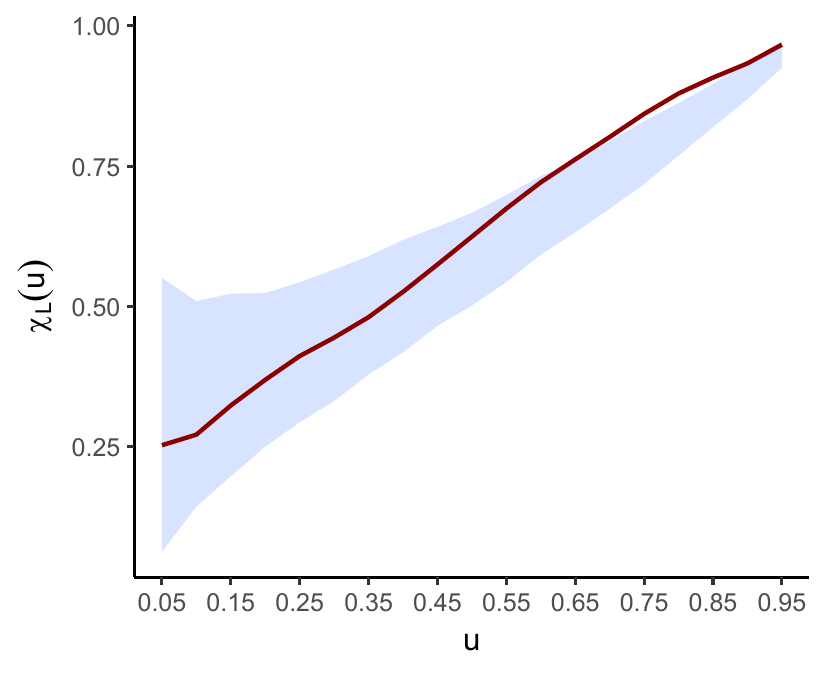}
            \caption{$\chi_{L}\left(u\right)$ envelope}
          \end{subfigure}
          \hfill
          \begin{subfigure}[b]{0.49\linewidth}
            \centering
            \includegraphics[width=\linewidth]{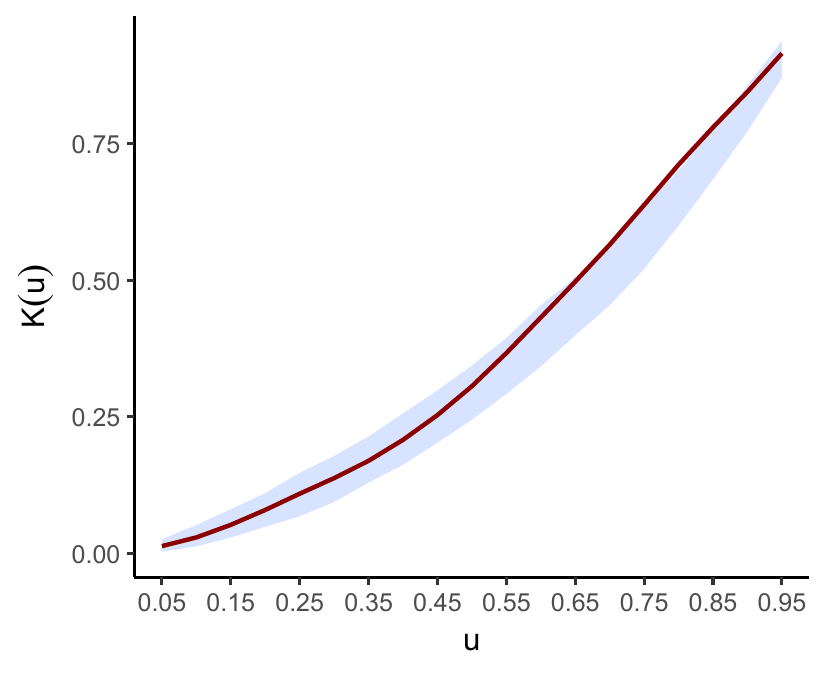}
            \caption{$K\left(u\right)$ envelope}
          \end{subfigure}
          \caption{Goodness-of-fit diagnostics for the Clayton copula model: The $\chi_L\left(u\right)$ function (left) probes tail association while the $K\left(u\right)$ function (right) assesses overall quadrant probabilities. Solid red curves represent the empirical statistics, while the blue ribbons show 95\% simulation envelopes. Curves remaining inside the ribbons indicate that the fitted model reproduces both tail behavior and global dependence adequately.}
          \label{fig:Clayton-gof-chiK}
        \end{figure}

        \afterpage{
        \begin{landscape}
        
          \begin{figure}[!ht]
            \centering
            \includegraphics{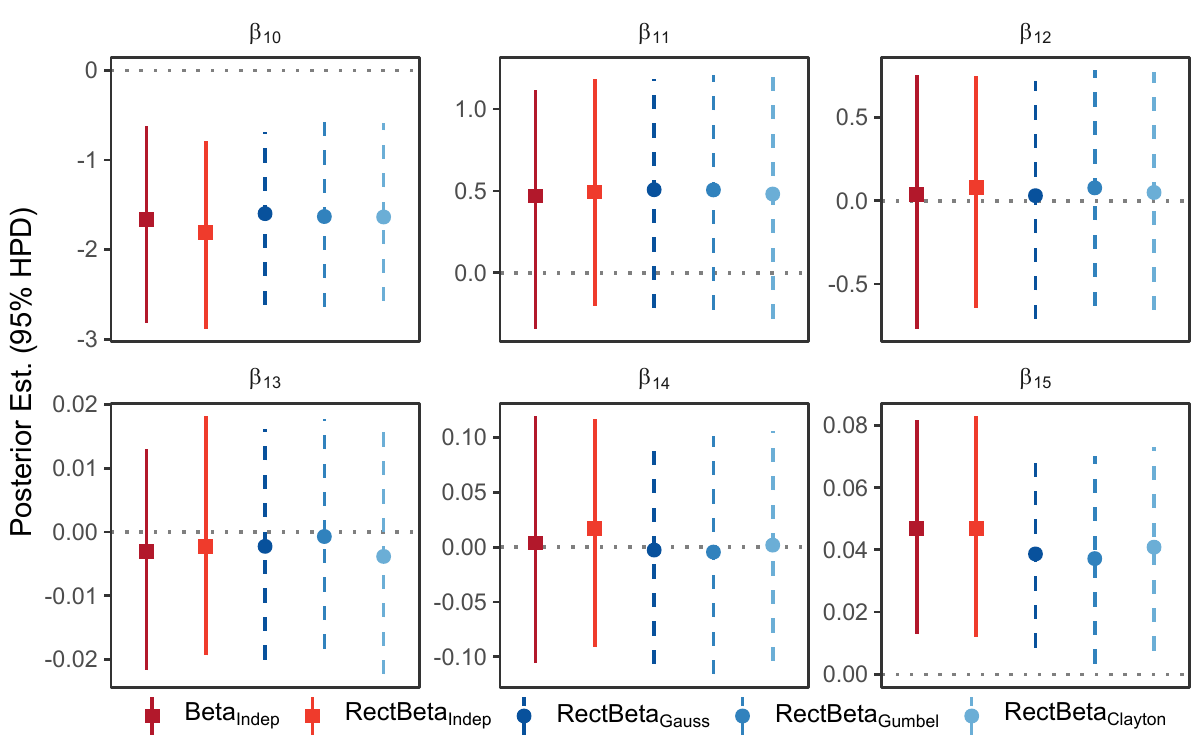}
            \caption{Posterior medians (symbols) and 95\% HPD intervals (vertical bars) for the six fixed-effect coefficients, 2003 outcome. Squares: independent-margin models; circles: copula models. Colors: dark red--light red for conventional vs. rectangular-beta independent models; dark--medium--light blue for Gaussian, Gumbel, and Clayton copulas, respectively. Solid lines denote independent margins, dashed lines denote copulas. The grey dotted horizontal line marks the zero-effect level; intervals that cross this line include zero in their 95\% HPD and, therefore, do not provide evidence of a non-zero effect. Coefficient definitions (design matrix order): $\beta_{10}$:~intercept; $\beta_{11}$:~$\text{AltMid}_{k\left(i\right)}$ -- indicator for mid- vs. high-elevation plot; $\beta_{12}$:~$\text{AspWest}_{k\left(i\right)}$ -- indicator for west-facing vs. east-facing slope; $\beta_{13}$:~$\text{Agr}_{1i}$ -- percent cover of \textit{Agrostis magellanica}; $\beta_{14}$:~$\text{LArea}_{1i}$ -- $\log\left(\text{cushion area}\right)$; $\beta_{15}$:~$\text{CO}_{1i}$ -- percent cover of all other vascular and non-vascular plants.}
            \label{fig:betaHPD-2003}
          \end{figure}

        \end{landscape}
    }

        \begin{landscape}
        
            \section{Tables} \label{sec:TAB_SUPP}
            
            \setlength{\tabcolsep}{0.15cm}
            \begin{ThreePartTable}
                \begin{TableNotes}[flushleft]\footnotesize
                        \item[a] $\Phi^{-1}\left(\cdot\right)$ is the quantile (inverse CDF) of the standard normal distribution, and $\Phi_{2,\theta}\left(x, y\right)$ is the bivariate normal CDF with zero means, unit variances, and correlation $\theta$. The Gaussian copula log-density is $\log\left(c_{\theta}\left(u, v\right)\right) = -\tfrac{1}{2} \log\left(1 - \theta^{2}\right) - \frac{z_{1}^{2} - 2\theta z_{1}z_{2} + z_{2}^{2}}{2\left(1 - \theta^{2}\right)} + \tfrac{1}{2} \left(z_{1}^{2} + z_{2}^{2}\right)$, where $z_{1} = \Phi^{-1}\left(u\right)$ and $z_{2} = \Phi^{-1}\left(v\right)$.
                        \item[b] Negative dependence ($\tau < 0$) is available only for the Gaussian family. When $\tau = 0$, the copula reduces to (strict) independence: $\theta = 0$ for Gaussian, $\theta = 1$ for Gumbel, and $\theta \to 0^{+}$ for Clayton (independence attained only in the limit).
                \end{TableNotes}
                \begin{longtable}{lllllll}
                        \caption{Closed-form copula CDFs $C_{\theta}\left(u, v\right)$ and their parameterization through Kendall's $\tau$. Tail-dependence coefficients $\left(\lambda_{L}, \lambda_{U}\right)$ appear in the last column.} \label{tab:C-theta-long}\\
                        \toprule
                        \textbf{Family} & \textbf{Tail} &
                        {$\bm{C_{\theta}\left(u, v\right)}$\textsuperscript{\bf a}} &
                        {$\bm{\theta\left(\tau\right)}$ \textbf{mapping}} &
                        {$\bm{\tau}$\textbf{-range}} &
                        {$\bm{\theta}$\textbf{-range}\textsuperscript{\bf b}} &
                        {$\bm{\left(\lambda_{L}, \lambda_{U}\right)}$} \\
                        \midrule
                        \endfirsthead
                        \multicolumn{7}{@{}l}{\itshape Table \thetable\ continued}\\
                        \toprule
                        \textbf{Family} & \textbf{Tail} &
                        {$\bm{C_{\theta}\left(u, v\right)}$\textsuperscript{\bf a}} &
                        {$\bm{\theta\left(\tau\right)}$ \textbf{mapping}} &
                        {$\bm{\tau}$\textbf{-range}} &
                        {$\bm{\theta}$\textbf{-range}\textsuperscript{\bf b}} &
                        {$\bm{\left(\lambda_{L}, \lambda_{U}\right)}$} \\
                        \midrule
                        \endhead
                        \insertTableNotes\\
                        \endlastfoot
                        Gaussian & Sym. &
                        $C_{\theta}\left(u, v\right) = \Phi_{2,\theta}\left(\Phi^{-1}\left(u\right), \Phi^{-1}\left(v\right)\right)$ &
                        $\theta = \sin\left(\tfrac{\pi}{2} \tau\right)$ &
                        $\left(-1, 1\right)$ & $\left(-1, 1\right)$ & $\left(0, 0\right)$ \\
                    
                        Gumbel & Upper &
                        $C_{\theta}\left(u, v\right) = \exp\left\{- \left[\left(-\log\left(u\right)\right)^{\theta} + \left(-\log\left(v\right)\right)^{\theta} \right]^{1 / \theta}\right\}$ &
                        $\theta = 1 / \left(1 - \tau\right)$ &
                        $\left(0, 1\right)$ & $\left[1, \infty\right)$ & $\left(0, 2 - 2^{1 / \theta}\right)$ \\
                        Clayton & Lower &
                        $C_{\theta}\left(u, v\right) = \left(u^{-\theta} + v^{-\theta} - 1\right)^{-1 / \theta}$ &
                        $\theta = 2\tau / \left(1 - \tau\right)$ &
                        $\left(0, 1\right)$ & $\left(0, \infty\right)$ & $\left(2^{-1 / \theta}, 0\right)$ \\
                        \bottomrule
                \end{longtable}
            \end{ThreePartTable}
        \end{landscape}

        \begin{landscape}

        \begin{center}
            \setlength\LTleft{0pt}
            \setlength\LTright{0pt}
        
            \setlength{\tabcolsep}{0.1cm}
            \begin{footnotesize}
                \begin{longtable}{l *{15}{S[table-format=3.3]}}
                    \caption{Posterior point estimates and 95\% HPD intervals.\label{tab:posterior}}\\
                    
                    \toprule
                    & \multicolumn{3}{c}{\textbf{Beta$_{\text{Indep}}$}} &
                      \multicolumn{3}{c}{\textbf{RectBeta$_{\text{Indep}}$}} &
                      \multicolumn{3}{c}{\textbf{RectBeta$_{\text{Gauss}}$}} &
                      \multicolumn{3}{c}{\textbf{RectBeta$_{\text{Gumbel}}$}} &
                      \multicolumn{3}{c}{\textbf{RectBeta$_{\text{Clayton}}$}} \\
                    \cmidrule(lr){2-4}\cmidrule(lr){5-7}\cmidrule(lr){8-10}\cmidrule(lr){11-13}\cmidrule(lr){14-16}
                    \textbf{Par.} &
                    \textbf{Est.} & \multicolumn{2}{c}{\textbf{95\% HPD}} &
                    \textbf{Est.} & \multicolumn{2}{c}{\textbf{95\% HPD}} &
                    \textbf{Est.} & \multicolumn{2}{c}{\textbf{95\% HPD}} &
                    \textbf{Est.} & \multicolumn{2}{c}{\textbf{95\% HPD}} &
                    \textbf{Est.} & \multicolumn{2}{c}{\textbf{95\% HPD}} \\
                    \midrule
                    \endfirsthead
                    
                    \multicolumn{16}{l}{\textit{(Continued from previous page)}}\\
                    \toprule
                    & \multicolumn{3}{c}{\textbf{Beta$_{\text{Indep}}$}} &
                      \multicolumn{3}{c}{\textbf{RectBeta$_{\text{Indep}}$}} &
                      \multicolumn{3}{c}{\textbf{RectBeta$_{\text{Gauss}}$}} &
                      \multicolumn{3}{c}{\textbf{RectBeta$_{\text{Gumbel}}$}} &
                      \multicolumn{3}{c}{\textbf{RectBeta$_{\text{Clayton}}$}} \\
                    \cmidrule(lr){2-4}\cmidrule(lr){5-7}\cmidrule(lr){8-10}\cmidrule(lr){11-13}\cmidrule(lr){14-16}
                    \textbf{Par.} &
                    \textbf{Est.} & \multicolumn{2}{c}{\textbf{95\% HPD}} &
                    \textbf{Est.} & \multicolumn{2}{c}{\textbf{95\% HPD}} &
                    \textbf{Est.} & \multicolumn{2}{c}{\textbf{95\% HPD}} &
                    \textbf{Est.} & \multicolumn{2}{c}{\textbf{95\% HPD}} &
                    \textbf{Est.} & \multicolumn{2}{c}{\textbf{95\% HPD}} \\
                    \midrule
                    \endhead
                    
                    \midrule
                    \multicolumn{16}{r}{\textit{(Continued on next page)}}\\
                    \endfoot
                    
                    \bottomrule
                    \endlastfoot
                    
                    $\beta_{10}$ & -1.661 & [-2.818; & -0.617] & -1.808 & [-2.883; & -0.788] & -1.599 & [-2.617; & -0.686] & -1.631 & [-2.639; & -0.531] & -1.636 & [-2.576; & -0.592] \\ 
                      $\beta_{11}$ & 0.469 & [-0.341; & 1.117] & 0.493 & [-0.201; & 1.185] & 0.507 & [-0.217; & 1.182] & 0.506 & [-0.224; & 1.207] & 0.481 & [-0.280; & 1.235] \\ 
                      $\beta_{12}$ & 0.040 & [-0.768; & 0.759] & 0.080 & [-0.646; & 0.748] & 0.031 & [-0.710; & 0.719] & 0.078 & [-0.631; & 0.784] & 0.050 & [-0.658; & 0.772] \\ 
                      $\beta_{13}$ & -0.003 & [-0.022; & 0.013] & -0.002 & [-0.019; & 0.018] & -0.002 & [-0.020; & 0.016] & -0.001 & [-0.018; & 0.018] & -0.004 & [-0.022; & 0.016] \\ 
                      $\beta_{14}$ & 0.004 & [-0.105; & 0.120] & 0.017 & [-0.091; & 0.116] & -0.003 & [-0.107; & 0.097] & -0.004 & [-0.116; & 0.101] & 0.002 & [-0.104; & 0.106] \\ 
                      $\beta_{15}$ & 0.047 & [0.013; & 0.082] & 0.047 & [0.012; & 0.083] & 0.039 & [0.009; & 0.072] & 0.037 & [0.003; & 0.070] & 0.041 & [0.008; & 0.073] \\ 
                      $\beta_{20}$ & 0.124 & [-0.949; & 1.089] & -0.595 & [-1.543; & 0.386] & -0.925 & [-1.862; & 0.003] & -0.853 & [-1.797; & 0.108] & -0.814 & [-1.770; & 0.192] \\ 
                      $\beta_{21}$ & 0.442 & [-0.128; & 0.967] & 0.530 & [0.005; & 1.093] & 0.544 & [0.082; & 1.041] & 0.559 & [0.193; & 0.967] & 0.527 & [0.040; & 1.017] \\ 
                      $\beta_{22}$ & 0.377 & [-0.207; & 0.896] & 0.453 & [-0.085; & 0.939] & 0.459 & [0.014; & 0.902] & 0.450 & [0.021; & 0.840] & 0.445 & [-0.038; & 0.989] \\ 
                      $\beta_{23}$ & 0.007 & [-0.009; & 0.023] & 0.007 & [-0.006; & 0.021] & 0.006 & [-0.007; & 0.019] & 0.005 & [-0.007; & 0.018] & 0.007 & [-0.006; & 0.021] \\ 
                      $\beta_{24}$ & -0.218 & [-0.336; & -0.101] & -0.139 & [-0.246; & -0.034] & -0.095 & [-0.206; & 0.010] & -0.101 & [-0.213; & 0.007] & -0.106 & [-0.223; & -0.001] \\ 
                      $\beta_{25}$ & -0.003 & [-0.037; & 0.031] & -0.000 & [-0.029; & 0.030] & -0.012 & [-0.041; & 0.016] & -0.016 & [-0.045; & 0.011] & -0.010 & [-0.039; & 0.017] \\ 
                      $\phi_{1}$ & 10.119 & [8.610; & 11.802] & 0.039 & [0.000; & 0.110] & 0.044 & [0.000; & 0.117] & 0.070 & [0.002; & 0.150] & 0.042 & [0.000; & 0.115] \\ 
                      $\phi_{2}$ & 8.641 & [7.363; & 10.022] & 0.153 & [0.063; & 0.241] & 0.170 & [0.083; & 0.259] & 0.199 & [0.113; & 0.291] & 0.151 & [0.065; & 0.236] \\ 
                      $\rho_{1}$ & & & & 11.142 & [9.023; & 13.487] & 11.146 & [9.002; & 13.405] & 11.324 & [9.128; & 13.702] & 10.938 & [8.958; & 13.418] \\ 
                      $\rho_{2}$ & & & & 14.924 & [11.677; & 18.137] & 15.762 & [12.612; & 19.025] & 16.107 & [13.072; & 19.985] & 15.324 & [12.274; & 18.643] \\ 
                      $\sigma_{1}$ & 0.410 & [0.158; & 0.863] & 0.403 & [0.162; & 0.892] & 0.390 & [0.140; & 0.819] & 0.389 & [0.148; & 0.857] & 0.411 & [0.164; & 0.860] \\ 
                      $\sigma_{2}$ & 0.285 & [0.057; & 0.674] & 0.245 & [0.056; & 0.631] & 0.225 & [0.046; & 0.538] & 0.205 & [0.014; & 0.481] & 0.248 & [0.069; & 0.597] \\ 
                      $\tau$ & & & & & & & 0.260 & [0.196; & 0.328] & 0.295 & [0.220; & 0.361] & 0.168 & [0.107; & 0.226] \\ 
                \end{longtable}
            \end{footnotesize}
        \end{center}

        \end{landscape}

        \afterpage{
            \setlength{\tabcolsep}{0.15cm}
            \begin{center}
                \begin{longtable}{
                  S[table-format = 1.2]
                  S[table-format = 1.2]
                  S[table-format = 1.2]
                  l
                  S[table-format = +1.4]
                  S[table-format = +1.4]
                  S[table-format = +1.4]
                  S[table-format = 1.3]
                  l
                  S[table-format = +1.4]
                  S[table-format = +1.4]
                  S[table-format = 1.3]
                }
                \caption{Simulation study: frequentist performance of the Gaussian-rectangular‑beta copula model. The column ``True" lists the data-generating value for each parameter. For sample sizes $n=300$ and $n=500$, the table reports the Monte Carlo mean bias, RMSE, and the coverage probability (CP) of the nominal 95\% HPD interval, each summarized over 200 simulated datasets.}
                \label{tab:bias_rmse_cov}
                \\
                \toprule
                \multicolumn{5}{c}{} &
                \multicolumn{3}{c}{$\bm{n=300}$} &
                \multicolumn{1}{c}{} &
                \multicolumn{3}{c}{$\bm{n=500}$} \\
                \cmidrule(lr){6-8} \cmidrule(lr){10-12}
                \textbf{\bm{$\phi_1$}} &
                \textbf{\bm{$\phi_2$}} &
                \textbf{\bm{$\tau$}} &
                \textbf{Param.} &
                \textbf{True} &
                \textbf{Bias} & \textbf{RMSE} & \textbf{CP} &
                \multicolumn{1}{c}{} &
                \textbf{Bias} & \textbf{RMSE} & \textbf{CP} \\
                \midrule
                \endfirsthead
                \multicolumn{12}{l}{\itshape Continued from previous page}\\
                \addlinespace[0.5ex]        
                \toprule
                \multicolumn{5}{c}{} &
                \multicolumn{3}{c}{$\bm{n=300}$} &
                \multicolumn{1}{c}{} &
                \multicolumn{3}{c}{$\bm{n=500}$} \\
                \cmidrule(lr){6-8} \cmidrule(lr){10-12}
                \textbf{\bm{$\phi_1$}} &
                \textbf{\bm{$\phi_2$}} &
                \textbf{\bm{$\tau$}} &
                \textbf{Param.} &
                \textbf{True} &
                \textbf{Bias} & \textbf{RMSE} & \textbf{CP} &
                \multicolumn{1}{c}{} &
                \textbf{Bias} & \textbf{RMSE} & \textbf{CP} \\
                \midrule
                \endhead
                
                \midrule
                \multicolumn{12}{r}{\itshape Continued on next page} \\
                \endfoot
                
                \bottomrule
                \endlastfoot
                    0.05 & 0.05 & 0 & $\bm{\beta}_{11}$ & -0.8473 & 0.0085 & 0.0345 & 0.940 &  & 0.0030 & 0.0242 & 0.980 \\* 
                       &  &  & $\bm{\beta}_{12}$ & 0.3 & -0.0004 & 0.0309 & 0.965 &  & -0.0025 & 0.0267 & 0.915 \\* 
                       &  &  & $\bm{\beta}_{21}$ & 0.4055 & -0.0024 & 0.0185 & 0.930 &  & -0.0012 & 0.0165 & 0.945 \\* 
                       &  &  & $\bm{\beta}_{22}$ & -0.3 & 0.0019 & 0.0282 & 0.945 &  & 0.0015 & 0.0216 & 0.975 \\* 
                       &  &  & $\bm{\phi}_1$ & 0.05 & 0.0104 & 0.0388 & 0.960 &  & 0.0040 & 0.0279 & 0.945 \\* 
                       &  &  & $\bm{\phi}_2$ & 0.05 & 0.0079 & 0.0208 & 0.940 &  & 0.0045 & 0.0265 & 0.955 \\* 
                       &  &  & $\bm{\rho}_1$ & 50 & 0.6735 & 5.0297 & 0.920 &  & 0.6145 & 3.6426 & 0.945 \\* 
                       &  &  & $\bm{\rho}_2$ & 50 & 0.5138 & 5.0394 & 0.940 &  & 0.2236 & 3.5275 & 0.980 \\* 
                       &  &  & $\bm{\tau}$ & 0 & -0.0016 & 0.0378 & 0.965 &  & -0.0004 & 0.0289 & 0.935 \\ 
                      0.05 & 0.05 & 0.25 & $\bm{\beta}_{11}$ & -0.8473 & 0.0050 & 0.0297 & 0.925 &  & 0.0024 & 0.0153 & 0.960 \\* 
                       &  &  & $\bm{\beta}_{12}$ & 0.3 & -0.0033 & 0.0321 & 0.940 &  & -0.0032 & 0.0234 & 0.940 \\* 
                       &  &  & $\bm{\beta}_{21}$ & 0.4055 & -0.0024 & 0.0179 & 0.950 &  & -0.0017 & 0.0135 & 0.960 \\* 
                       &  &  & $\bm{\beta}_{22}$ & -0.3 & -0.0006 & 0.0291 & 0.945 &  & -0.0017 & 0.0221 & 0.950 \\* 
                       &  &  & $\bm{\phi}_1$ & 0.05 & 0.0075 & 0.0290 & 0.980 &  & 0.0064 & 0.0182 & 0.940 \\* 
                       &  &  & $\bm{\phi}_2$ & 0.05 & 0.0062 & 0.0202 & 0.940 &  & 0.0033 & 0.0142 & 0.970 \\* 
                       &  &  & $\bm{\rho}_1$ & 50 & 0.3238 & 4.9512 & 0.920 &  & 0.6227 & 3.5860 & 0.965 \\* 
                       &  &  & $\bm{\rho}_2$ & 50 & 0.6758 & 4.6351 & 0.965 &  & 0.0868 & 3.6907 & 0.940 \\* 
                       &  &  & $\bm{\tau}$ & 0.25 & 0.0023 & 0.0319 & 0.950 &  & 0.0020 & 0.0271 & 0.945 \\ 
                      0.2 & 0.05 & 0 & $\bm{\beta}_{11}$ & -0.8473 & 0.0025 & 0.0360 & 0.925 &  & 0.0059 & 0.0362 & 0.950 \\* 
                       &  &  & $\bm{\beta}_{12}$ & 0.3 & -0.0036 & 0.0328 & 0.955 &  & -0.0039 & 0.0291 & 0.950 \\* 
                       &  &  & $\bm{\beta}_{21}$ & 0.4055 & -0.0021 & 0.0182 & 0.950 &  & 0.0000 & 0.0135 & 0.965 \\* 
                       &  &  & $\bm{\beta}_{22}$ & -0.3 & 0.0028 & 0.0281 & 0.960 &  & 0.0005 & 0.0234 & 0.945 \\* 
                       &  &  & $\bm{\phi}_1$ & 0.2 & 0.0032 & 0.0398 & 0.965 &  & 0.0031 & 0.0382 & 0.940 \\* 
                       &  &  & $\bm{\phi}_2$ & 0.05 & 0.0095 & 0.0213 & 0.960 &  & 0.0032 & 0.0154 & 0.930 \\* 
                       &  &  & $\bm{\rho}_1$ & 50 & 0.9360 & 6.0114 & 0.925 &  & -0.0292 & 4.2994 & 0.930 \\* 
                       &  &  & $\bm{\rho}_2$ & 50 & 0.6542 & 4.9711 & 0.945 &  & 0.3162 & 3.5080 & 0.945 \\* 
                       &  &  & $\bm{\tau}$ & 0 & -0.0016 & 0.0385 & 0.915 &  & -0.0005 & 0.0283 & 0.950 \\ 
                      0.2 & 0.05 & 0.25 & $\bm{\beta}_{11}$ & -0.8473 & 0.0061 & 0.0460 & 0.945 &  & 0.0022 & 0.0235 & 0.940 \\* 
                       &  &  & $\bm{\beta}_{12}$ & 0.3 & -0.0044 & 0.0330 & 0.965 &  & -0.0029 & 0.0237 & 0.965 \\* 
                       &  &  & $\bm{\beta}_{21}$ & 0.4055 & -0.0016 & 0.0175 & 0.945 &  & -0.0010 & 0.0132 & 0.965 \\* 
                       &  &  & $\bm{\beta}_{22}$ & -0.3 & -0.0021 & 0.0297 & 0.935 &  & 0.0015 & 0.0235 & 0.945 \\* 
                       &  &  & $\bm{\phi}_1$ & 0.2 & 0.0079 & 0.0469 & 0.950 &  & 0.0027 & 0.0278 & 0.945 \\* 
                       &  &  & $\bm{\phi}_2$ & 0.05 & 0.0067 & 0.0197 & 0.935 &  & 0.0028 & 0.0141 & 0.935 \\* 
                       &  &  & $\bm{\rho}_1$ & 50 & 0.4880 & 5.6596 & 0.925 &  & 0.3360 & 3.7530 & 0.955 \\* 
                       &  &  & $\bm{\rho}_2$ & 50 & 0.2313 & 4.8209 & 0.955 &  & -0.1295 & 3.6055 & 0.950 \\* 
                       &  &  & $\bm{\tau}$ & 0.25 & 0.0035 & 0.0326 & 0.970 &  & 0.0019 & 0.0243 & 0.975 \\ 
                      0.2 & 0.2 & 0 & $\bm{\beta}_{11}$ & -0.8473 & 0.0017 & 0.0291 & 0.960 &  & 0.0008 & 0.0222 & 0.975 \\* 
                       &  &  & $\bm{\beta}_{12}$ & 0.3 & -0.0002 & 0.0333 & 0.940 &  & 0.0018 & 0.0261 & 0.955 \\* 
                       &  &  & $\bm{\beta}_{21}$ & 0.4055 & -0.0007 & 0.0225 & 0.930 &  & -0.0002 & 0.0170 & 0.945 \\* 
                       &  &  & $\bm{\beta}_{22}$ & -0.3 & 0.0002 & 0.0273 & 0.990 &  & 0.0028 & 0.0210 & 0.970 \\* 
                       &  &  & $\bm{\phi}_1$ & 0.2 & 0.0046 & 0.0376 & 0.940 &  & -0.0002 & 0.0286 & 0.945 \\* 
                       &  &  & $\bm{\phi}_2$ & 0.2 & 0.0040 & 0.0352 & 0.935 &  & 0.0031 & 0.0268 & 0.930 \\* 
                       &  &  & $\bm{\rho}_1$ & 50 & 0.6764 & 5.2870 & 0.970 &  & -0.2265 & 3.9018 & 0.935 \\* 
                       &  &  & $\bm{\rho}_2$ & 50 & 0.9426 & 5.7887 & 0.940 &  & 0.5601 & 4.4220 & 0.970 \\* 
                       &  &  & $\bm{\tau}$ & 0 & -0.0012 & 0.0381 & 0.940 &  & 0.0019 & 0.0292 & 0.940 \\ 
                      0.2 & 0.2 & 0.25 & $\bm{\beta}_{11}$ & -0.8473 & 0.0016 & 0.0359 & 0.920 &  & -0.0003 & 0.0324 & 0.950 \\* 
                       &  &  & $\bm{\beta}_{12}$ & 0.3 & -0.0012 & 0.0349 & 0.930 &  & -0.0004 & 0.0255 & 0.945 \\* 
                       &  &  & $\bm{\beta}_{21}$ & 0.4055 & -0.0035 & 0.0193 & 0.940 &  & -0.0006 & 0.0156 & 0.950 \\* 
                       &  &  & $\bm{\beta}_{22}$ & -0.3 & 0.0007 & 0.0289 & 0.965 &  & -0.0002 & 0.0251 & 0.940 \\* 
                       &  &  & $\bm{\phi}_1$ & 0.2 & 0.0039 & 0.0417 & 0.935 &  & 0.0021 & 0.0343 & 0.940 \\* 
                       &  &  & $\bm{\phi}_2$ & 0.2 & 0.0060 & 0.0332 & 0.945 &  & 0.0030 & 0.0261 & 0.950 \\* 
                       &  &  & $\bm{\rho}_1$ & 50 & 0.7086 & 5.7143 & 0.930 &  & -0.0215 & 4.4903 & 0.935 \\* 
                       &  &  & $\bm{\rho}_2$ & 50 & 0.9612 & 6.3191 & 0.915 &  & 0.9422 & 4.1166 & 0.960 \\* 
                       &  &  & $\bm{\tau}$ & 0.25 & -0.0025 & 0.0334 & 0.955 &  & -0.0016 & 0.0250 & 0.950 \\ 
                \end{longtable}
            \end{center}
        }

    \end{appendices}
    
\end{document}